%% file: paper.tex
\documentclass[11pt,a4paper]{article}

\usepackage[utf8]{inputenc}  
\usepackage[german,english]{babel}
\usepackage{latexsym,amsbsy,amssymb,amsmath,amsfonts}
\usepackage{epsfig}
\usepackage{euscript}
\usepackage{units}
\usepackage{fullpage}
\usepackage{pdflscape}
\usepackage{indent}
\usepackage{enumitem}
\usepackage{rotating}
\usepackage{xcolor}

\usepackage{rotating}
\usepackage{tikz}
\usetikzlibrary{shapes}
\usetikzlibrary{automata}
\usetikzlibrary{arrows}
\usetikzlibrary{calc}
\usepackage[colorlinks,linkcolor={red!75!black},citecolor={blue!75!black},urlcolor={blue!80!black}]{hyperref}

\input{comphead007}

\begin{document}
\selectlanguage{english}

\def\sqsubsetneq{\mathrel{\sqsubseteq\kern-0.92em\raise-0.15em\hbox{\rotatebox{313}{\scalebox{1.1}[0.75]{\(\shortmid\)}}}\scalebox{0.3}[1]{\ }}}
\def\sqsupsetneq{\mathrel{\sqsupseteq\kern-0.92em\raise-0.15em\hbox{\rotatebox{313}{\scalebox{1.1}[0.75]{\(\shortmid\)}}}\scalebox{0.3}[1]{\ }}}

\newcommand{\io}{\mathrm{io\tn{-}}}

\newcommand{\ioeq}{\mathop{\mbox{$\mathop{=}\limits^{\mathrm{\scriptscriptstyle io}}$}}}

\newcommand{\notioeq}{\mathop{\mbox{$\mathop{=}\limits^{\mathrm{\scriptscriptstyle \mr i\ml .\ml o\ml .}}\!\!\!\!\!\raisebox{0.3ex}[0ex][0ex]{$/$}\mr\mr\mr$}}}

\newcommand{\ioin}{\mathop{\mbox{$\mathop{\in}\limits^{\mbox{\raisebox{-0.2ex}[0ex][0ex]{$\mathrm{\scriptscriptstyle \mr\mr i\ml .\ml o\ml .}$}}}$}}}

\newcommand{\notioin}{\mathop{\mbox{$\mathop{\in}\limits^{\mbox{\raisebox{-0.2ex}[0ex][0ex]{$\mathrm{\scriptscriptstyle \mr\mr i\ml .\ml o\ml .}$}}}\!\!\!\!\!\ml\raisebox{0.35ex}[0ex][0ex]{$/$}\mr\mr\mr$}}}

\newcommand{\iosub}{\mathop{\mbox{$\mathop{\subseteq}\limits^{\mbox{\raisebox{-0.2ex}[0ex][0ex]{$\mathrm{\scriptscriptstyle \mr\mr i\ml .\ml o\ml .}$}}}$}}}

\newcommand{\notiosub}{\mathop{\mbox{$\mathop{\subseteq}\limits^{\mbox{\raisebox{-0.2ex}[0ex][0ex]{$\mathrm{\scriptscriptstyle \mr\mr i\ml .\ml o\ml .}$}}}\!\!\!\!\!\ml\raisebox{0.35ex}[0ex][0ex]{$/$}\mr\mr\mr$}}}

\newcommand{\aeeq}{\mathop{\mbox{$\mathop{=}\limits^{\mathrm{\scriptscriptstyle \mr a\ml .\ml e\ml .}}$}}}

\newcommand{\notaeeq}{\mathop{\mbox{$\mathop{=}\limits^{\mathrm{\scriptscriptstyle \mr a\ml .\ml e\ml .}}\!\!\!\!\!\raisebox{0.3ex}[0ex][0ex]{$/$}\mr\mr\mr$}}}

\newcommand{\aein}{\mathop{\mbox{$\mathop{\in}\limits^{\mbox{\raisebox{-0.2ex}[0ex][0ex]{$\mathrm{\scriptscriptstyle \mr\mr a\ml .\ml e\ml .}$}}}$}}}

\newcommand{\notaein}{\mathop{\mbox{$\mathop{\in}\limits^{\mbox{\raisebox{-0.2ex}[0ex][0ex]{$\mathrm{\scriptscriptstyle \mr\mr a\ml .\ml e\ml .}$}}}\!\!\!\!\!\ml\raisebox{0.35ex}[0ex][0ex]{$/$}\mr\mr\mr$}}}

\newcommand{\aesub}{\mathop{\mbox{$\mathop{\subseteq}\limits^{\mbox{\raisebox{-0.2ex}[0ex][0ex]{$\mathrm{\scriptscriptstyle \mr\mr a\ml .\ml e\ml .}$}}}$}}}

\newcommand{\notaesub}{\mathop{\mbox{$\mathop{\subseteq}\limits^{\mbox{\raisebox{-0.2ex}[0ex][0ex]{$\mathrm{\scriptscriptstyle \mr\mr a\ml .\ml e\ml .}$}}}\!\!\!\!\!\ml\raisebox{0.35ex}[0ex][0ex]{$/$}\mr\mr\mr$}}}

\newcommand{\dosescomment}[1]{}

\newcommand{\psim}[1][]{\le^{\mathrm{p}#1}}

\newcommand{\NPCm}{\tn{NPC}^{\tn{p}}_{\tn{m}}}
\newcommand{\NPCO}{\tn{NPC}^{\tn{p,}O}_{\tn{m}}}
\newcommand{\NPCT}{\tn{NPC}^{\tn{p}}_{\tn{T}}}
\newcommand{\NPCmpoly}{\tn{NPC}^{\tn{p/poly}}_{\tn{m}}}
\newcommand{\NPCmiopoly}{\tn{NPC}^{\tn{io-p/poly}}_{\tn{m}}}
\newcommand{\NPCtwott}{{\rm NPC}^{\rm p}_{\text{\rm 2-tt}}}
\newcommand{\NPCli}{\text{\rm NPC}_{\rm m, li}^{\rm p}}
\renewcommand{\Pol}{\tn{Pol}}
\newcommand{\Pad}{\tn{Pad}}
\newcommand{\Hom}{\tn{Hom}}
\newcommand{\Hm}{\tn{H}^{\tn{p}}_{\tn{m}}}
\newcommand{\Hmli}{\tn{H}^{\tn{p}}_{\tn{m,li}}}
\newcommand{\Hone}{\tn{H}^{\tn{p}}_{\tn{1}}}
\newcommand{\HT}{\tn{H}^{\tn{p}}_{\tn{T}}}
\newcommand{\HsnT}{\tn{H}^{\tn{p}}_{\tn{snT}}}
\newcommand{\Hmpoly}{\tn{H}^{\tn{p/poly}}_{\tn{m}}}
\newcommand{\Hmiopoly}{\tn{H}^{\tn{io\tn{-}p/poly}}_{\tn{m}}}

\renewcommand{\P}{\mathbb{P}}
\newcommand{\Pone}{\P_1}
\newcommand{\Pthree}{\P_3}
\newcommand{\Podd}{\P^{\ge 3}}

\newcommand{\ran}{\tn{ran}}

\newcommand{\DisjcoNP}{\mathrm{DisjCoNP}}

\title{An Oracle Separating Conjectures about\\Incompleteness in the Finite Domain\footnote{This work was
        supported by a grant of the Studienstiftung des deutschen Volkes (German Academic Scholarship Foundation).}}

\author{Titus Dose\\Ostbayerische Technische Hochschule Regensburg}
\maketitle

\begin{abstract}\noindent
    Pudl{\'a}k \cite{pud17} lists several major complexity theoretic conjectures
    relevant to proof complexity and asks for oracles that separate pairs of corresponding
    relativized conjectures. Among these conjectures are:
    \begin{itemize}
        \item $\mathsf{DisjNP}$: The class of all disjoint NP-pairs does not have many-one complete problems.
        \item $\mathsf{SAT}$: NP does not contain many-one complete sets that have P-optimal proof systems.
        \item $\mathsf{UP}$: UP does not have many-one complete problems.
        \item $\mathsf{NP}\cap\mathsf{coNP}$: $\cNP\cap\ccoNP$ does not have many-one complete problems.
    \end{itemize}

    As one answer to this question, we construct an oracle
    relative to which (among others) $\mathsf{DisjNP}$, $\neg \mathsf{SAT}$, $\mathsf{UP}$, and
    $\mathsf{NP}\cap\mathsf{coNP}$ hold. Hence, there is no relativizable proof for the implication
    $\mathsf{DisjNP}\wedge \mathsf{UP}\wedge
        \mathsf{NP}\cap\mathsf{coNP}\Rightarrow\mathsf{SAT}$.
    In particular, regarding the conjectures by Pudl{\'a}k this extends a result by Khaniki \cite{kha19}.
\end{abstract}


\section{Introduction}
The main motivation for the present paper is an article by Pudlák \cite{pud17}
that is ``motivated by the problem of finding finite versions of classical
incompleteness theorems'', investigates major complexity theoretic conjectures relevant
to proof complexity, discusses their relations, and in particular draws new
connections between the conjectures.
Among others, Pudl{\'a}k conjectures the following assertions (note that within the present
paper all reductions are polynomial-time-bounded):
\begin{itemize}
    \item   $\mathsf{CON}$ (resp., $\mathsf{SAT}$): $\ccoNP$ (resp., $\cNP$) does not contain many-one complete
          sets that have P-optimal proof systems. Equivalently: TAUT (resp., SAT),
          the set of all propositional tautologies (resp., all satisfiable propositional formulas), does
          not have P-optimal proof systems \cite{kmt03}.
    \item   $\mathsf{CON}^{\mathsf{N}}$: $\ccoNP$ does not contain many-one complete
          sets that have optimal proof systems. Equivalently: TAUT does not have optimal proof systems
          \cite{kmt03}.\\
          (note that $\mathsf{CON}^{\mathsf{N}}$ is the non-uniform version of
          $\mathsf{CON}$)
    \item   $\mathsf{DisjNP}$ (resp., $\mathsf{DisjCoNP}$): The class of all disjoint $\cNP$-pairs
          (resp., $\ccoNP$-pairs) does not have many-one complete elements.
    \item   $\mathsf{TFNP}$: The class of all total polynomial search problems (more commonly
          known as total NP search problems) does not have complete elements.
    \item   $\mathsf{NP}\cap\mathsf{coNP}$ (resp., $\mathsf{UP}$): $\cNP\cap\ccoNP$ (resp., $\cUP$,
          the class of problems accepted by $\cNP$ machines with at most one accepting path
          for each input) does not have many-one complete elements.
\end{itemize}

The reason why Pudlák lays emphasis on this selection
of conjectures is that these conjectures can be
equivalently formulated as statements about unprovability
of particular $\Pi_1$ sentences. Pudlák's investigation is
overall driven by the fundamental question of the connection
between logical strength of theories and computational complexity
\cite{pud17}.

The following figure contains the uniform conjectures by Pudlák and illustrates the state of the art
regarding (i) known implications and (ii) separations
in terms of oracles that prove the non-existence of
relativizable proofs for implications.
\newcommand{\CCON}{\mathsf{CON}}
\newcommand{\CCONN}{\mathsf{CON}^\mathsf{N}}
\newcommand{\CDisjNP}{\mathsf{DisjNP}}
\newcommand{\CUP}{\mathsf{UP}}
\newcommand{\CRFN}{\mathsf{RFN}_1}
\newcommand{\CPNP}{\mathsf{P}\ne\mathsf{NP}}
\newcommand{\CNPcoNP}{\mathsf{NP}\cap\mathsf{coNP}}
\newcommand{\CSAT}{\mathsf{SAT}}
\newcommand{\CCONSAT}{\CCON\vee\CSAT}
\newcommand{\CTFNP}{\mathsf{TFNP}}
\newcommand{\CDisjCoNP}{\mathsf{DisjCoNP}}

\begin{center}
    \begin{figure}[ht]
        \begin{tikzpicture}[->,>=stealth',initial text={},shorten >=1pt,auto,node distance=2cm,
                thin,main node/.style={draw=none,font=\sffamily\Large}, thin]

            \node[main node] (1)  {$\CDisjNP$};
            \phantom{\node[main node] (2) [below of=1] {$\CCONN$};}
            \node[main node] (3) [right of=2] {$\CUP$};
            \node[main node] (4) [ below of=3] {$\CCON$};
            \node[main node] (12) [right of=4]{};
            \node[ main node] (14) [ below of=4] {};
            \node[ main node] (15) [ right of=14] {};
            \node[ main node] (5) [ right of=15] {$\CCON\vee\CSAT$};
            \node[ main node] (6) [ below  of=5] {$\cP\ne\cNP$};
            \node[ main node] (7) [ right of=12] {$\CNPcoNP$};
            \node[main node] (13) [right of=7]{};
            \node[ main node] (8) [ right of=13] {$\CSAT$};
            \node[ main node] (9) [ above of=8] {$\CTFNP$};
            \node[ main node] (10) [ above of=9] {$\CDisjCoNP$};

            \path[every node/.style={font=\sffamily\small, color  = black}]
            (1) edge[] node[left] {} (4)
            (3) edge node [] {} (4)
            (4) edge node [] {} (5)
            (5) edge node [] {} (6)
            (7) edge node [] {} (5)
            (8) edge node [] {} (5)
            (9) edge node [] {} (8)
            (10) edge [] node  {} (9);

            \path[every node/.style={font=\sffamily\small, color  = black},dashed]
            (1) edge node [above,sloped] {\cite{dg19}} (3)
            (7) edge node [above] {\phantom{.....}\cite{dos19b}} (4)
            (1) edge [bend left] node [sloped,above] {Cor~\ref{cor_0147120472134544}} (8)
            (3) edge [bend left] node [sloped,above] {Cor~\ref{cor_0147120472134544}\phantom{........}} (8)
            (7) edge [] node [above] {Cor~\ref{cor_0147120472134544}} (8)
            (10) edge [] node [above,sloped] {\cite{kha19}\phantom{...............}} (4)
            (4) edge[bend left] node[left]{\cite{gssz04}}(1)
            (6) edge[bend right=60] node[right] {\cite{dos19a}}(5);
        \end{tikzpicture}
        \caption{\label{fig_1047120471}
            Solid arrows mean implications. All implications occurring in
            the graphic have relativizable proofs (relativizable proofs for $\CDisjNP\Rightarrow \CCON$,
            $\CUP\Rightarrow \CCON$, and $\CNPcoNP\Rightarrow \CCON\vee\CSAT$ can be found in \cite{kmt03};
            relativizable proofs  for $\CDisjCoNP\Rightarrow \CTFNP$ and
            $\CTFNP\Rightarrow\CSAT$ can be found in \cite{pud17,bkm09}; the other implications
            are trivial).
            A dashed arrow from one conjecture $\mathsf{A}$ to another conjecture $\mathsf{B}$ means that
            there is an oracle $X$ against the implication $\mathsf{A}\Rightarrow
                \mathsf{B}$, which means $\mathsf{A}\wedge\neg \mathsf{B}$ relative to $X$.
            Cor~\ref{cor_0147120472134544} refers to the oracle constructed in the present paper.
            \newline Pudl{\'a}k \cite{pud17} also defines the conjecture $\CRFN$ and lists it between $\CCON\vee\CSAT$ and
            $\CPNP$, i.e., $\CCON\vee\CSAT\Rightarrow \CRFN\Rightarrow \CPNP$. Khaniki \cite{kha19} even shows $\CCON\vee\CSAT\Leftrightarrow \CRFN$, which is why we omit $\CRFN$ in the figure.
            For a definition of $\CRFN$ we refer to \cite{pud17}.
        }
    \end{figure}
\end{center}

The main conjectures of \cite{pud17} are $\mathsf{CON}$ and $\mathsf{TFNP}$. Let us give some
background on these conjectures (for details we refer to \cite{pud13})
and on the notion of disjoint pairs. The first
main conjecture $\mathsf{CON}$ refers to the notion of proof systems introduced
by Cook and Reckhow \cite{cr79}, who define a proof system for a set $A$ to be a total,
polynomial-time computable function whose range is $A$.

The subsequent paragraph is due to \cite{dg19} and explains a logical characterization
of $\mathsf{CON}$ and $\mathsf{CON}^\mathsf{N}$.
$\mathsf{CON}$ has an interesting connection to some finite version of an incompleteness
statement.
Denote by $\tn{Con}_T(n)$ the finite consistency of a finitely axiomatized theory $T$, i.e.,
$\tn{Con}_T(n)$ is the statement that $T$ does not have proofs of contradiction of length $\le n$.
Kraj{\'{\i}}{\v c}ek and Pudl{\'a}k \cite{kp89} raise
the conjectures $\mathsf{CON}$ and $\mathsf{CON}^{\mathsf{N}}$ and show that the latter is
equivalent to the statement that
there is no finitely axiomatized theory $S$
which proves the finite consistency $\tn{Con}_T(n)$
for every finitely axiomatized theory $T$
by a proof of polynomial length in $n$. 
In other words, $\neg \mathsf{CON}^{\mathsf{N}}$
expresses that a weak version of Hilbert's program
(to prove the consistency of all mathematical theories) is possible \cite{pud96}.
Correspondingly, $\neg\mathsf{CON}$ is equivalent to the existence of a theory $S$ such that,
for each fixed finitely axiomatized
theory $T$, proofs of $\tn{Con}_T(n)$ in $S$ can be constructed in polynomial time in $n$
\cite{kp89}.

The conjecture $\mathsf{TFNP}$,
raised by Megiddo and Papadimitriou \cite{mp91}, is implied by
the non-existence of disjoint coNP-pairs \cite{bkm09,pud17}, and implies that no
NP-complete set has P-optimal proof systems \cite{bkm09,pud17}. It states the non-existence of
total polynomial search problems that are complete with respect to polynomial reductions,
where a total polynomial search problem (i) is represented by a polynomial $p$
and a binary relation $R$ $\boldsymbol{\in\cP}$ satisfying $\forall_x\exists_y\; |y|\le p(|x|)\wedge (x,y)\in R$ and
(ii) is the following computational task: On input $x$ compute some $y$ with $|y|\le p(|x|)\wedge
    (x,y)\in R$. In other words, total polynomial search problems are represented by nondeterministic
polynomial-time multivalued functions with values that are polynomially verifiable and guaranteed to exist
\cite{mp91}.

The notion of disjoint NP-pairs, i.e., pairs $(A,B)$ with $A\cap B = \emptyset$
and $A,B\in\cNP$, has its origin in public-key cryptography and
characterizes promise problems \cite{ey80,esy84,gs88}.
Razborov \cite{raz94} connects disjoint pairs with the concept of propositional proof systems (pps),
i.e., proof systems for the set of propositional tautologies TAUT, defining
for each pps $f$ a disjoint NP-pair, the so-called canonical pair of $f$, where
the canonical pair of an optimal pps $f$ is complete \cite{raz94,kmt03}. Hence, putting it another way,
$\mathsf{DisjNP}\Rightarrow \mathsf{CON}^{\mathsf{N}}$.

In contrast to the many implications Pudlák \cite{pud17} only knows of one oracle separating
two of the relativized conjectures, namely an oracle by Glaßer et al.\ \cite{gssz04} relative to
which $\CDisjNP\wedge\neg\CCONN$. Therefore, Pudlák asks for further oracles
showing relativized conjectures to be different.

Khaniki \cite{kha19} partially answers this question: besides showing
two of the conjectures to be equivalent he presents two oracles $\calV$ and $\calW$
showing that $\mathsf{SAT}$ and $\mathsf{CON}$ (as well as
$\mathsf{TFNP}$ and $\mathsf{CON}$)
are independent in relativized worlds which means that none of the two
possible implications between the two conjectures has a relativizable proof.
To be more precise,
relative to $\calV$, there exist P-optimal
propositional proof systems but no many-one complete
disjoint coNP-pairs, where ---as mentioned above--- the latter
implies $\mathsf{TFNP}$ and $\mathsf{SAT}$ \cite{pud17,bkm09}.
Relative to $\calW$, there exist no optimal propositional proof
systems, but each total polynomial search problem has a polynomial-time solution, where
the latter implies $\neg\mathsf{SAT}$ \cite{km00}.

Dose and Gla\ss{}er \cite{dg19} construct an oracle $X$ that also separates
some of the above relativized conjectures. Relative to $X$ there exist no
many-one complete disjoint $\cNP$-pairs, $\cUP$ has many-one complete problems,
and $\cNP\cap\ccoNP$ does not have many-one complete problems.
In particular, relative to $X$, there do not exist P-optimal propositional
proof systems. Thus, among others, $X$ shows that the conjectures
$\mathsf{CON}$ and $\mathsf{UP}$ cannot be proven equivalent with relativizable proofs.

In two further papers \cite{dos19a,dos19b}, the author adds two more oracle
to this list proving that there is no relativizable
proof for the implications (i) $\cP\ne\cNP\Rightarrow \CCONSAT$ and (ii)
$\CNPcoNP\Rightarrow \CCON$. Formally,
relative to the first oracle, $\cP\ne\cNP$ and all non-empty problems in $\cNP$ or $\ccoNP$
have $\cP$-optimal proof systems and relative to the second oracle, $\cNP\cap\ccoNP$ does not have
complete problems and all non-empty sets in $\ccoNP$ have P-optimal proof systems.

\paragraph*{Our Contribution}
In the present paper we construct an oracle $O$ relative to which
\begin{enumerate}
    \item The class of all disjoint NP-pairs does not have many-one complete elements.
    \item Each (equivalently some) many-one complete set for $\cNP$ has P-optimal proof systems.
    \item UP does not contain many-one complete problems.
    \item $\cNP\cap\ccoNP$ does not contain many-one complete problems.
\end{enumerate}
Indeed, relative to $O$, there even exist no disjoint $\cNP$-pairs that are hard for $\cUP\cap\ccoUP$,
which implies both 1 and 4. Figure~\ref{fig_1047120471} illustrates that $O$ yields one
of the strongest oracle results that Pudlák \cite{pud17} asks for since $\mathsf{DisjNP}$, $\mathsf{UP}$,
and $\mathsf{NP}\cap\mathsf{coNP}$ are the strongest conjectures in their respective branches
in Figure~\ref{fig_1047120471}
whereas $\mathsf{SAT}$ is the weakest conjecture that is not relativizably implied
by the three other conjectures. In other words, in Figure~\ref{fig_1047120471}
all conjectures on the left hold and all others are wrong relative to the oracle,
in particular, for all conjectures in Figure~\ref{fig_1047120471} we know whether
they hold relative to $O$.

Among others, the oracle shows that there are no relativizable
proofs for the implications $\CNPcoNP\Rightarrow \CSAT$ and $\CUP\Rightarrow \CSAT$. Let us now focus
on the properties 1 and 2 of the oracle. Regarding these,
our oracle has similar properties as the aforementioned oracle $\calW$ by
Khaniki \cite{kha19}: both oracles show that there is no relativizable proof for
the implication $\CCONN\Rightarrow \CSAT$. Relative to Khaniki's oracle
$\calW$ it even holds that each total polynomial search
problem has a polynomial time solution, which implies not only $\neg\mathsf{SAT}$
but also that all optimal proof systems for SAT (such as the standard proof system for SAT)
are P-optimal \cite{km00}. Regarding
Pudlák's conjectures, however, our oracle $O$ extends Khaniki's result as relative
to $O$ we have
the even stronger result that there is no relativizable proof for the implication
$\mathsf{DisjNP}\Rightarrow \mathsf{SAT}$.
Since due to the oracle $\calV$ by Khaniki \cite{kha19} none of the
implications $\mathsf{DisjCoNP}\Rightarrow \mathsf{DisjNP}$,
$\mathsf{TFNP}\Rightarrow \mathsf{DisjNP}$, and
$\mathsf{SAT}\Rightarrow \mathsf{DisjNP}$
can be proven relativizably,
our oracle shows that $\mathsf{DisjNP}$ is independent
of each of the conjectures
$\mathsf{DisjCoNP}$, $\mathsf{TFNP}$, and $\mathsf{SAT}$
in relativized worlds, which means that none of the six possible implications
has a relativizable proof.

Let us emphasize the fact that the conjectures branch (confer
Figure~\ref{fig_1047120471}): Pudlák's main open problem
is to find a ``general conjecture about incompleteness and computational complexity''
from which the current conjectures follow as special cases \cite{pud17}.
The aforementioned oracle $\calV$ by Khaniki \cite{kha19} and our oracle suggest
that none of the current conjectures is such a general conjecture: the two oracles prove that for
each conjecture $\mathsf{A}$ in Figure~\ref{fig_1047120471} there is one conjecture $\mathsf{B}$
that is not implied by $\mathsf{A}$ relative to all oracles. Hence, showing that one of the
current conjectures solves Pudlák's main open problem requires non-relativizable proofs.

\paragraph*{Open Questions}
There have been recently constructed three oracles by Fabian Egidy, Anton Ehrmanntraut, and Christian Glaßer
    [private communication, unpublished] relative to which (i) $\CDisjNP\wedge\neg(\CNPcoNP)$,
(ii) $\CUP\wedge\neg(\CNPcoNP)$, and (iii) $\CDisjCoNP\wedge\neg(\CNPcoNP)$, respectively.
Hence, if three more oracles with the following properties
are constructed, then for each pair $(\mathsf{A},\mathsf{B})$ of the conjectures in Figure~\ref{fig_1047120471}
we either know a relativizable proof for the implication $\mathsf{A}\Rightarrow\mathsf{B}$ or we know an oracle relative to which
$\mathsf{A}\wedge \neg\mathsf{B}$:
\begin{enumerate}
    \item $\CUP\wedge\neg\CDisjNP$ (or better $\CUP\wedge\neg\CCONN$)
    \item $\CSAT\wedge\neg\CTFNP$
    \item $\CTFNP\wedge\neg \CDisjCoNP$.
\end{enumerate}
Therefore, we ask for oracles relative to which 1, 2, or 3 holds.

\section{Preliminaries} \label{sec_prelim}
Most parts of this section originate from a previous paper \cite{dg19}.
Throughout this paper let $\Sigma$ be the alphabet $\{0,1\}$.
We denote the length of a word
$w\in\sow$ by $|w|$.
Let $\az^{\prec n} = \{w \in \sow ~|~ |w| \prec n\}$ for $\prec\in
    \{\le,<,=,>,\ge\}$, where in case $\az^{=n}$ we simply write $\az^n$.
The empty word is denoted by $\varepsilon$ and
the $i$-th letter of a word $w$ for $0 \le i < |w|$ is denoted by $w(i)$, i.e.,
$w = w(0) w(1) \cdots w(|w|-1)$.
If $v$ is a prefix of $w$, i.e., $|v|\le|w|$ and $v(i) = w(i)$
for all $0\le i <|v|$, then we write $v \sqsubseteq w$ or $w\sqsupseteq v$. If $v\sqsubseteq w$
and $|v| < |w|$, then we write $v\sqsubsetneq w$ or $w\sqsupsetneq v$.
For each finite set $Y \subseteq \sow$, let
$\ell(Y) = \sum_{w \in Y} |w|$.

Given two sets
$A$ and $B$, $A-B$ denotes the set difference
between $A$ and $B$, i.e., $A-B = \{a\in A\mid a\notin B\}$.
The complement of a set $A$ relative to the universe $U$
is denoted by $\oli{A}= U-A$. The universe will always be
apparent from the context.  Furthermore, the symmetric difference
is denoted by $\triangle$, i.e., $A\triangle B = (A-B)\cup (B-A)$
for arbitrary sets $A$ and $B$. For a finite set $A$ we denote
the cardinality of $A$ by $|A|$.

$\mathbb{Z}$ denotes the set of integers,
$\N$ denotes the set of natural numbers, and $\N^+ = \N-\{0\}$.
The set of primes is denoted by $\mathbb{P} = \{2,3,5,7,11,\ldots\}$
and $\Podd$ denotes the set $\mathbb{P}-\{2\}$.

We identify $\sow$ with $\N$
via the polynomial-time computable, polynomial-time invertible bijection
$w \mapsto \sum_{i<|w|} (1+w(i)) 2^{|w|-1-i}$,
which is a variant of the dyadic encoding.
Hence, notations, relations, and operations for $\sow$
are transferred to $\N$ and vice versa.
In particular, $|n|$ denotes the length of $n \in \N$.
We eliminate the ambiguity of the expressions $0^i$ and $1^i$
by always interpreting them over $\sow$.

Let $\langle \cdot \rangle : \bigcup_{i \ge 0} \N^i \rightarrow \N$
be an injective, polynomial-time computable, polynomial-time invertible
pairing function such that
$|\pairing{u_1, \ldots, u_n}| = 2(|u_1| + \cdots + |u_n| + n)$.

The domain and range of a function $t$ are denoted by
$\tn{dom}(t)$ and $\ran(t)$, respectively.

$\FP$, $\cP$, and $\cNP$ denote standard complexity classes \cite{pap94}.
Define $\opstyle{co\mathcal{C}} = \{A\subseteq\Sigma^*\mid \oli{A} \in\mathcal{C}\}$
for a class $\mathcal{C}$. $\cUP$ is the class of all problems accepted by
nondeterministic polynomial-time Turing machines that on each input have at most
one accepting path.
If $A,B\in\cNP$ and $A\cap B=\emptyset$, then we call $(A,B)$ a disjoint $\cNP$-pair.
The set of all disjoint $\cNP$-pairs is denoted by $\DisjNP$.
We also consider all these complexity classes in the presence of
an oracle $D$ and denote the corresponding classes by $\FP^D$, $\cP^D$, $\cNP^D$, and so on.

Let $M$ be an oracle Turing machine. $M^D(x)$ denotes the
computation of $M$ on input $x$ with $D$ as an oracle. For an
arbitrary oracle $D$ we let $L(M^D) = \{ x ~|~ M^D(x) \tn{ accepts}\}$,
where as usual in case $M$ is nondeterministic, the computation $M^D(x)$
accepts if and only if it has at least one accepting path.

For a deterministic polynomial-time oracle Turing transducer (i.e., a Turing machine computing
a function), depending on the context,
$F^D(x)$ either denotes the computation of $F$ on input $x$ with $D$ as an oracle
or the output of this computation.

\begin{definition}\label{definition_019470123712}
    A sequence $(M_i)_{i\in\N^+}$ is called a {\em standard enumeration} of
    nondeterministic, polynomial-time oracle Turing machines
    if it has the following properties:
    \begin{enumerate}
        \item All $M_i$ are nondeterministic, polynomial-time oracle Turing machines.
        \item For all oracles $D$ and all inputs $x$
              the computation $M_i^D(x)$ stops within $|x|^i + i$ steps.
        \item For every nondeterministic, polynomial-time oracle Turing machine $M$
              there exist infinitely many $i \in \N$ such that
              for all oracles $D$ it holds $L(M^D) = L(M_i^D)$.
        \item There exists a nondeterministic, polynomial-time oracle Turing machine $M$
              such that for all oracles $D$ and all inputs $x$ it holds that $M^D(\langle i,0^{|x|^i+i},x \rangle)$
              nondeterministically simulates the computation $M_i^D(x)$.
    \end{enumerate}
    Analogously we define standard enumerations of
    deterministic, polynomial-time (oracle) Turing transducers.
\end{definition}
Throughout this paper, we fix some standard enumerations.
Let $M_1,M_2,\dots$ be a standard enumeration of nondeterministic
polynomial-time oracle Turing machines. Then for every oracle $D$, the sequence
$(M_i)_{i\in\N^+}$ represents an enumeration of the languages in $\cNP^D$, i.e.,
$\cNP^D = \{L(M_i^D)\mid i\in\N^+\}$.
Let $F_1,F_2,\dots$ be a standard enumeration of polynomial time oracle
Turing transducers and $T_1,T_2,\dots$ be a standard enumeration of polynomial time
Turing transducers (i.e., not having access to an oracle).

In the present article we only use polynomial-time-bounded many-one reductions.
Let $D$ be an oracle. For problems $A,B\subseteq\sow$ we write $A\redm B$ (resp.,
$A\redm[,D] B$) if there exists an $f\in\FP$ (resp., $f\in\FP^D$) with
$\forall_{x\in\az^*} x\in A\Leftrightarrow f(x)\in B$. In this case we say that $A$
is polynomially many-one reducible to $B$ (relative to $D$).
A problem $A\subset\sow$ is $\redm$-hard (resp.,
$\redm[,D]$-hard) for a complexity class $\calC$ if $B\redm A$ (resp., $B\redm[,D] A$) for
all $B\in\calC$. Moreover, $A$ is called $\redm$-complete (resp., $\redm[,D]$-complete) for
a class $\calC$, if $A\in\calC$ and $A$ is $\redm$-hard (resp., $\redm[,D]$-hard) for $\calC$.

Now let $A,B,A',B'\subseteq\sow$
such that $A\cap B = A'\cap B' = \emptyset$.
In this paper we always use the following reducibility for disjoint pairs
\cite{raz94}. $(A',B')$ is polynomially many-one
reducible to $(A,B)$ relative to $D$, denoted by $(A',B')\redmprom[,D](A,B)$,
if there exists an $f\in\FP^D$ with $f(A')\subseteq A$ and $f(B')\subseteq B$.
If $A' = \oli{B'}$, then we also write $A'\redm[,D](A,B)$ instead of
$(A',B')\redmprom[,D] (A,B)$.

We say that $(A,B)$ is {\em $\redmprom[,D]$-hard} ({\em $\redmprom[,D]$-complete}) {\em for} $\DisjNP^D$
if $(A',B')\redmprom[,D] (A,B)$ for all $(A',B')\in \DisjNP^D$ (and $(A,B)\in \DisjNP^D$).
Moreover, a pair $(A,B)$ is {\em $\redm[,D]$-hard for} $\cUP^D\cap\ccoUP^D$ if
$A'\redm[,D] (A,B)$ for every $A'\in\cUP^D\cap\ccoUP^D$.

By the properties of standard enumerations, for each oracle $D$ the problem
$$K^D = \{\pairing{0^i,0^t,x}\mid i,t,x\in\N, i>0, \tn{ and $M_i^D(x)$ accepts within $t$ steps}\}$$
is $\redm[,D]$-complete for $\cNP^D$ (in particular it is in $\cNP^D$).

\begin{definition}[\cite{cr79}]
    A function $f \in \FP$ is called a {\em proof system} for the set $\ran(f)$.
    If $f(x) = y$, then we say that $x$ is an $f$-{\it proof} for $y$.
    For $f,g \in \FP$ we say that {\em $f$ is simulated by $g$} (resp.,
        {\em $f$ is $\cP$-simulated by $g$}) denoted by $f\le g$ (resp., $f \psim g$),
    if there exists a function $\pi$ (resp., a function $\pi \in \FP$)
    and a polynomial $p$ such that $|\pi(x)|\le p(|x|)$ and $g(\pi(x)) = f(x)$ for all $x$.
    A function $g \in \FP$ is {\em optimal} (resp., {\em $\cP$-optimal}),
    if $f\le g$ (resp., $f \psim g$) for all $f \in \FP$ with $\tn{ran(f)}=\tn{ran(g)}$.
    Corresponding relativized notions are obtained by using
    $\cP^D$, $\FP^D$, and $\psim[,D]$ in the definitions above.
\end{definition}The following proposition states the relativized version of a result by
K{\"o}bler, Messner, and Tor\'an \cite{kmt03},
which they show with a relativizable proof.
\begin{proposition}[\cite{kmt03}] \label{propo_pps_oracle}
    For every oracle $D$,
    if $A$ has a $\cP^D$-optimal (resp., optimal) proof system
    and $\emptyset\ne B \redm[,D]\! A$,
    then $B$ has a $\cP^D$-optimal (resp., optimal) proof system.
\end{proposition}

\begin{corollary} \label{coro_pps_oracle}
    For every oracle $D$,
    if there exists a $\redm[,D]\!$-complete $A \in \cNP^D$
    that has a $\cP^D$-optimal (resp., optimal) proof system,
    then all non-empty sets in $\cNP^D$
    have $\cP^D$-optimal (resp., optimal) proof systems.
\end{corollary}

\medskip
Let us introduce some notations that are designed
for the construction of oracles \cite{dg19}.
The support $\tn{supp}(t)$ of a real-valued function $t$
is the subset of the domain that
consists of all arguments that $t$ does not map to 0.
We say that a partial function $t$ is injective on its support if
$t(i,j) = t(i',j')$ for $(i,j),(i',j') \in \tn{supp}(t)$ implies $(i,j) = (i',j')$.
If a partial function $t$ is not defined at point $x$,
then $t \cup \{x \mapsto y\}$ denotes the continuation $t'$
of $t$ that at $x$ has value $y$ and satisfies
$\tn{dom}(t') = \tn{dom}(t)\cup\{x\}$

If $A$ is a set, then $A(x)$ denotes the characteristic function at point $x$,
i.e., $A(x)$ is $1$ if $x \in A$, and $0$ otherwise.
An oracle $D \subseteq \N$ is identified with its characteristic sequence
$D(0) D(1) \cdots$, which is an $\omega$-word.
In this way, $D(i)$ denotes both, the characteristic function at point $i$ and
the $i$-th letter of the characteristic sequence, which are the same.
A finite word $w$ describes an oracle that is partially defined, i.e.,
only defined for natural numbers $x<|w|$.
Occasionally, we use $w$ instead of the set $\{i \mid w(i)=1 \}$ and
write for example $A = w \cup B$, where $A$ and $B$ are sets (however, we do not
use the notation $|w|$ referring to the cardinality of the set $\{i \mid w(i)=1 \}$,
but only referring to the length of the word $w$).
In particular, for an oracle Turing machine $M$, the notation
$M^w(x)$ refers to $M^{\{i \mid w(i)=1 \}}(x)$ (hence, oracle queries that $w$ is not defined for
are answered by ``no''). Using $w$ instead of $\{i \mid w(i)=1 \}$
additionally allows us to define the following notion:
for a nondeterministic oracle Turing machine $M$, the computation $M^w(x)$ {\em definitely accepts}
if it contains a path that accepts and all queries
on this path are $<|w|$.
The computation $M^w(x)$ {\em definitely rejects}
if all paths reject and all queries are $<|w|$.
We say
that the computation $M^w(x)$ {\em is definite}
if it definitely accepts or definitely rejects.
Similarly, for a deterministic oracle Turing transducer $F$, the computation $F^w(x)$ is {\em definite}
if all its queries are $<|w|$.

\bigskip
Before we start with the construction of the oracle, we mention two standard results about $\cUP$-complete sets
and hard pairs.

\begin{proposition}\label{prop:queryfree}
    Let $O$ be an oracle.
    \begin{enumerate}
        \item $\cUP^O$ has $\redm[,O]$-complete sets if and only if $\cUP^O$ has $\redm$-complete sets.
        \item $\DisjNP^O$ has pairs that are $\redm[,O]$-hard for $\cUP^O\cap\ccoUP^O$ if and only if $\DisjNP^O$ has pairs that are $\redm$-hard for $\cUP^O\cap\ccoUP^O$.
    \end{enumerate}
\end{proposition}
\begin{proof}
    The ``if'' directions are trivial since any reduction that does not query the oracle is also a valid reduction relative to $O$.
    We now show the ``only if'' directions.

    Consider 1. Let $U$ be a $\redm[,O]$-complete set for $\cUP^O$. We define a new language $U' = \{ \pairing{i, 0^{p_i(|x|)}, x} \mid F_i^O(x) \in U \}$.
    Since $F_i^O$ is a deterministic polynomial-time reduction, a nondeterministic oracle machine for $U'$ with oracle $O$ that first deterministically
    computes $y = F_i^O(x)$ and then simulates the
    $\cUP^O$-machine for $U$ on $y$ has at most one accepting path for every input. Thus, $U' \in \cUP^O$.

    Moreover, every language $A \in \cUP^O$ reduces to $U$ via some oracle transducer $F_j^O$. Therefore, $A$ reduces to $U'$ via the mapping $x \mapsto \pairing{j, 0^{p_j(|x|)}, x}$.
    This reduction makes no oracle queries. Hence, $U'$ is $\redm$-complete for $\cUP^O$.

    Consider 2. Let $(A, B) \in \DisjNP^O$ be $\redm[,O]$-hard for $\cUP^O \cap \ccoUP^O$ and define the pair $(A', B')$ by
    \begin{align*}
        A' & = \{ \pairing{i, 0^{p_i(|x|)}, x} \mid F_i^O(x) \in A \}, \\
        B' & = \{ \pairing{i, 0^{p_i(|x|)}, x} \mid F_i^O(x) \in B \}.
    \end{align*}
    Since $A, B \in \cNP^O$ and each $F_i^O$ is a deterministic polynomial-time reduction, we have $A', B' \in \cNP^O$. Moreover, $A \cap B = \emptyset$ implies $A' \cap B' = \emptyset$, so $(A', B') \in \DisjNP^O$.
    Since $(A, B)$ is $\redm[,O]$-hard for $\cUP^O \cap \ccoUP^O$, for every language $L \in \cUP^O \cap \ccoUP^O$ there exists some oracle transducer $F_j^O$ such that $F_j^O(L) \subseteq A$ and $F_j^O(\oli{L}) \subseteq B$.
    Therefore, the mapping $x \mapsto \pairing{j, 0^{p_j(|x|)}, x}$ satisfies $x \in L \Rightarrow \pairing{j, 0^{p_j(|x|)}, x} \in A'$ and $x \in \oli{L} \Rightarrow \pairing{j, 0^{p_j(|x|)}, x} \in B'$.
    Since this mapping makes no oracle queries, it holds that $L \redm (A', B')$.
    Hence, $(A', B')$ is $\redm$-hard for $\cUP^O \cap \ccoUP^O$.
\end{proof}

\section{Oracle Construction}
\begin{lemma}[\cite{dg19}]\label{84189612339821692813}
    For all $y\le |w|$ and all $v\sqsupseteq w$ it holds $(y\in K^v \Leftrightarrow y\in K^w)$.
\end{lemma}
\begin{proof} We may assume $y = \pairing{0^i,0^t,x}$ for suitable $i\in\N^+$ and $t,x\in\N$,
    since otherwise, $y\notin K^w$ and $y\notin K^v$.
    For each $q$ that is queried within the first $t$ steps
    of $M_i^w(x)$ or $M_i^v(x)$ it holds that
    $|q| \le t < |y|$ and thus, $q < y$.
    Hence, these queries are answered the same way relative to $w$ and $v$,
    showing that $M_i^w(x)$ accepts within $t$ steps if and only if $M_i^v(x)$ accepts
    within $t$ steps.
\end{proof}

In the proof of the following theorem, we achieve an optimal proof system in
a way similar to \cite[Theorem 7.1]{dg19}. Moreover, the outer structure of the proof
follows \cite{dg19}.
\begin{theorem}\label{theorem_0917240914}
    There exists an oracle $O$ such that
    the following statements hold:
    \begin{itemize}
        \item $\DisjNP^O$ does not contain pairs that are $\redm[,O]$-hard
              for $\cUP^O\cap\ccoUP^O$.
        \item Each non-empty $L\in\cNP^O$ has $\cP^O$-optimal proof systems.
        \item $\cUP^O$ does not contain $\redm[,O]$-complete problems.
    \end{itemize}
\end{theorem}
The following Corollary immediately follows from
Theorem~\ref{theorem_0917240914}.
\begin{corollary}\label{cor_0147120472134544} There exists an oracle $O$ such
    that the following statements hold:
    \begin{itemize}
        \item $\DisjNP^O$ does not contain $\redmprom[,O]$-complete pairs.
        \item Each non-empty $L\in\cNP^O$ has $\cP^O$-optimal proof systems.
        \item $\cUP^O$ does not contain $\redm[,O]$-complete problems.
        \item $\cNP^O\cap\ccoNP^O$ does not contain $\redm[,O]$-complete problems.
    \end{itemize}
\end{corollary}
\begin{proof}{\bf of Theorem~\ref{theorem_0917240914}}
    Let $D$ be a (possibly partial) oracle, $p\in\Pthree$, and $q\in\Pone$, where
    $\Pthree = \P\cap \{4k+3\mid k\in\N\}$ and $\Pone = \P\cap\{4k+1\mid k\in\N\}$
    (note that $\Pthree$ and $\Pone$ are both infinite sets).
    We define
    \begin{eqnarray*}
        A_p^D &:=& \{ {0}^{p^k} \mid k\in\N^+, \exists_{x\in
            \az^{p^k}} x\in D\tn{ and $x$ odd}\} \cup \oli{
            \{ {0}^{p^k}\mid k\in\N^+\}}\\
        B_p^D &:=& \{ {0}^{p^k} \mid k\in\N^+, \exists_{x\in
            \az^{p^k}} x\in D\tn{ and $x$ even}\}\\
        C_q^D &:=& \{ {0}^{q^k}\mid k\in\N^+, \exists_{x\in
            \az^{q^k}} x\in D\}
    \end{eqnarray*}
    Note that $A_p^D,B_p^D\in\cNP^D$ and $A_p^D = \oli{
            B_p^D}$ if $|\az^{p^k}\cap D| = 1$ for each $k\in\N^+$.
    In that case $A_p^D\in\cUP^D\cap\ccoUP^D$.
    Moreover, $C_q^D\in\cUP^D$ if $|\az^{q^k}\cap D|\le 1$
    for each $k\in\N^+$.

    Note that throughout this proof we sometimes omit the oracles in
    the superscript, e.g., we write $\cNP$ or
    $A_p$ instead of $\cNP^D$ or $A_p^D$. However,
    we do not do that in the ``actual'' proof but only when
    explaining ideas in a loose way
    in order to give the reader the intuition behind the
    occasionally very technical arguments.

    \bigskip For $i\in\N^+$ and $x,y\in\N$
    we write $c(i,x,y) := \pairing{0^i,x, 0^{|x|^i + i}, 0^{|x|^i + i},y,y}$.
    Note that $|c(i,x,y)|$ is even, polynomially bounded in $|i| + |x| + |y|$, and by the properties
    of the pairing function $\pairing{\cdot}$,
    \begin{align}\label{eq_140712041003}
        \forall_{i\in\N^+, x,y\in\N}\;|c(i,x,y)| > 4\cdot \max( |x|^i + i,|y|).
    \end{align}\begin{claim}[\cite{dg19}] \label{claim_7378194873}
        Let $w\in\az^*$ be an oracle, $i\in\N^+$, and $x\in\N$. If
        $c(i,x,y)\le |w|$ for some $y\in\N$, then the following holds.
        \begin{enumerate}
            \item $F_i^w(x)$ is definite and $F_i^w(x)<|w|$.
            \item $(F_i^w(x)\in K^w\Leftrightarrow F_i^w(x)\in K^v)$
                  for all $v \sqsupseteq w$.
        \end{enumerate}
    \end{claim}
    \begin{proof}
        As the running time of $F_i^w(x)$ is bounded by
        $|x|^i + i < |c(i,x,y)| < c(i,x,y) \le |w|$,
        the computation $F_i^w(x)$ is definite and its output
        is less than $|w|$. Hence, 1 holds.
        Consider 2.
        It suffices to show that $K^v(q) = K^w(q)$ for all $q < |w|$ and all $v\sqsupseteq w$.
        This holds by Lemma~\ref{84189612339821692813}.
    \end{proof}

    \paragraph*{Preview of the construction} We sketch some of the very basic ideas our construction uses.
    \begin{enumerate}[wide, labelwidth=!, labelindent=0pt,topsep=0pt]
        \item For all positive $i\ne j$
              the construction tries to achieve that $(L(M_i),L(M_j))$ is not a disjoint $\cNP$-pair.
              If this is not possible, then $(L(M_i), L(M_j))$ inherently
              is a disjoint NP-pair. Once we know this, we choose some prime $p\in\Pthree$,
              ensure $A_p=\oli{B_p}$ and $A_p\in\cUP\cap\ccoUP$ in the further construction,
              and diagonalize against all $\FP$-functions such that
              $A_p$ is not $\redm$-reducible to $(L(M_i),L(M_j))$.
        \item For all $i\ge 1$ the construction intends to make sure
              that $F_i$ is not a proof system for $K$. If this is not possible, then
              $F_i$ inherently is a proof system for $K$. Then we start to encode
              the values of $F_i$ into the oracle, which will allow us to define a
              proof system that simulates all others relative to the oracle. However,
              it is important to also allow encodings for functions that are not known
              to be proof systems for $K$ yet. Thus, the final oracle will also contain
              encodings for functions that are not proof systems for $K$.
        \item For all $i\ge 1$ the construction tries to ensure that $M_i$ is not a $\cUP$-machine.
              In case this is impossible, we know that $M_i$ inherently is a $\cUP$-machine, which enables us
              to diagonalize against all $\FP$-functions making sure that $C_q$ for some $q\in\Pone$
              that we choose is not reducible to $L(M_i)$.
    \end{enumerate}

    During the construction we maintain a growing collection
    of requirements that is specified by a partial function belonging to the set
    $$\calT = \Big\{t:\N^+\cup (\N^+)^2\to\parbox[t]{130mm}{$\Z\mid
    \tn{dom}(t)$ is finite, $t$ is injective on its support,
        \begin{itemize}[itemsep=0pt,topsep=0pt]
        \item $t(\N^+)\subseteq \{0\}\cup\N^+$
        \item $t(\{(i,i)\mid i\in\N^+\}) \subseteq \{0\}\cup \{-q\mid q\in\Pone\}$
        \item $t(\{(i,j)\in (\N^+)^2\mid i\ne j\})\subseteq \{0\}
    \cup\{-p\mid p\in \Pthree\}$\Big\}.
        \end{itemize}}$$

    A partial oracle $w\in\Sigma^*$ is called $t$-valid for $t\in\calT$
    if it satisfies the following requirements.
    \begin{itemize}
        \item[V1]   For all $i\in\N^+$ and all $x,y\in\N$, if $c(i,x,y)\in w$,
              then $F_i^w(x) = y\in K^w$.\\
              (meaning: if the oracle contains the codeword $c(i,x,y)$,
              then $F_i^w(x)$ outputs $y$ and $y \in K^w$;
              hence, $c(i,x,y) \in w$ is a proof for $y \in K^w$.)
        \item[V2]   For all distinct $i,j\in\N^+$, if $t(i,j) = 0$, then there exists $x$
              such that $M_i^w(x)$ and $M_j^w(x)$ definitely accept.\\
              (meaning: for every extension of the oracle, $(L(M_i),L(M_j))$
              is not a disjoint NP-pair.)
        \item[V3]   For all distinct $i,j\in\N^+$ with $t(i,j) = -p$ for some $p\in\Pthree$
              and each $k\in\N^+$, it holds (i) $|\az^{p^k}\cap w|\le 1$ and
              (ii) if $w$ is defined for all words of length $p^k$, then
              $|\az^{p^k}\cap w| = 1$.
              \\(meaning: if $t(i,j) =-p$, then ensure
              $A_p\in\cUP\cap\ccoUP$ relative to the final oracle.)
        \item[V4]   For all $i\in\N^+$ with $t(i) = 0$, there exists $x$ such that $F_i^w(x)$ is
              definite and $F_i^w(x)\notin K^v$ for all $v\sqsupseteq w$.\\
              (meaning: for every extension of the oracle, $F_i$ is not a proof system for $K$)
        \item[V5]   For all $i\in\N^+$ and $x\in\N$ with $0 \!<\! t(i)\!\le\! c(i,x,F_i^w(x)) \!<\! |w|$,
              it holds $c(i,x,F_i^w(x)) \in w$.\\
              (meaning: if $t(i) > 0$, then from $t(i)$ on, we encode
              the values of $F_i$ into the oracle.\\
              Note that V5 is not in contradiction to V3 or V7 as $|c(\cdot,\cdot,\cdot)|$ is even.)
        \item[V6]   For all $i\in\N^+$ with $t(i,i) = 0$, there exists $x$ such that $M_i^w(x)$ is definite
              and has two accepting paths.
              \\(meaning: for every extension of the oracle, $M_i$ is not a $\cUP$-machine.)
        \item[V7]   For all $i\in\N^+$ with $t(i,i) = -q\in\Pone$ and each $k\in\N^+$, it holds
              $|\az^{q^k}\cap w| \le 1$.
              \\(meaning: if $t(i,i) = -q$, ensure that $C_q$ is in $\cUP$.)
    \end{itemize}

    The subsequent claim follows directly from the definition of $t$-valid.
    \begin{claim}\label{claim_90124780576}
        Let $t,t'\in\calT$ such that $t'$ is an extension of $t$.
        For all oracles $w\in\az^*$, if $w$ is $t'$-valid, then $w$ is $t$-valid.
    \end{claim}
    \begin{claim}\label{claim_sandwich}
        Let $t\in\calT$ and $u,v,w\in\az^*$ be oracles such that $u\sqsubseteq v\sqsubseteq w$
        and both $u$ and $w$ are $t$-valid. Then $v$ is $t$-valid.
    \end{claim}
    \begin{proof}
        $v$ satisfies V2, V4, and V6 since $u$ satisfies these conditions. Moreover, $v$ satisfies
        V3 and V7 as $w$ satisfies these conditions.

        Let $i\in\N^+$ and $x,y\in\N$ such that
        $c(i,x,y)\in v$. Then $c(i,x,y)\in w$ and as $w$ is $t$-valid, we obtain by V1 that
        $F_i^w(x) = y \in K^w$. Claim~\ref{claim_7378194873} yields that $F_i^v(x)$ is definite
        and $F_i^v(x) \in K^v\Leftrightarrow F_i^v(x)\in K^w$. This yields
        that $F_i^v(x) = F_i^w(x) = y$ and $K^v(y) = K^w(y) = 1$. Thus, $v$ satisfies V1.

        Now let $i\in\N^+$ and $x\in\N$ such that $0<t(i)\le c(i,x,F_i^v(x))<|v|$. Again,
        by Claim~\ref{claim_7378194873}, $F_i^v(x)$ is definite and thus, $F_i^v(x) = F_i^w(x)$.
        As $|v| \le |w|$ and $w$ is $t$-valid, we obtain by V5 that $c(i,x,F_i^v(x)) =
            c(i,x,F_i^w(x))\in w$. Since $v\sqsubseteq w$ and $|v|>c(i,x,F_i^v(x))$, we obtain
        $c(i,x,F_i^v(x))\in v$, which shows that $v$ satisfies V5.
    \end{proof}

    \paragraph*{Oracle construction} Let $T$ be an enumeration of $\bigcup_{i=1}^3 (\N^+)^i$
    having the property that $(i,j)$ appears
    earlier than $(i,j,r)$ for all $i,j,r\in\N^+$.
    Each element of $\bigcup_{i=1}^3 (\N^+)^i$
    is considered as a task. We treat the tasks in the order specified by $T$ and after
    treating a task we remove it and possibly other tasks from $T$ and continue with the next task
    that still is in $T$ (i.e., the first task of the current $T$). We start
    with the nowhere defined function $t_0$ and the $t_0$-valid oracle $w_0 =
        \varepsilon$. Then we begin treating the tasks. For each task we choose an extension of the current function from $\calT$
    and an extension of the oracle. This way we define functions $t_1,t_2,\dots$ in $\calT$ such that $t_{i+1}$
    is an extension of $t_i$ and partial oracles $w_0\sqsubsetneq w_1\sqsubsetneq w_2
        \sqsubsetneq\dots$ such that each $w_i$ is $t_i$-valid. So for each task we strictly extend
    the oracle and are allowed to add more requirements (by extending the valid function)
    that have to be maintained in the further construction. Finally, we choose
    $O = \bigcup_{i = 0}^\infty w_i$ (note that $O$ is totally defined since in
    each step we strictly extend the oracle).
    We now describe step $s>0$, which starts with some $t_{s-1}\in\calT$ and a $t_{s-1}$-valid oracle $w_{s-1}$
    and treats the first task that still is in $T$ choosing an extension $t_s\in\calT$
    of $t_{s-1}$ and a $t_s$-valid $w_s\sqsupsetneq w_{s-1}$ (it will be
    argued later that all these steps are indeed possible).
    Let us recall that each task is immediately deleted from $T$ after it is treated.
    We study five cases depending on the form of the task that is treated in step~$s$.

    \begin{itemize}
        \item task $i$: Let $t' = t_{s-1}\cup \{i\mapsto 0\}$. If there
              exists a $t'$-valid partial oracle $v\sqsupsetneq w_{s-1}$,
              then let $t_s = t'$ and $w_s$ be the least $t'$-valid partial oracle $\sqsupsetneq w_{s-1}$.
              Otherwise, let $t_s = t_{s-1}\cup \{i \mapsto |w_{s-1}|\}$ (note that this implies that $w_{s-1}$
              is $t_s$-valid) and choose $w_s = w_{s-1}b$ with $b\in\{0,1\}$
              such that $w_s$ is $t_s$-valid.
              \\(meaning: try to ensure that $F_i$ is not a proof system for $K$. If this is impossible,
              require that from now on the values of $F_i$ are encoded into the oracle.)
        \item task $(i,j)$ with $i\ne j$: Let $t' =
                  t_{s-1}\cup \{(i,j)\mapsto 0\}$. If there
              exists a $t'$-valid partial oracle $v\sqsupsetneq w_{s-1}$,
              then let $t_s = t'$, define $w_s$ to be the least $t'$-valid partial oracle $\sqsupsetneq w_{s-1}$,
              and delete all tasks $(i,j,\cdot)$ from $T$.
              Otherwise, let $z = |w_{s-1}|$, choose some $p\in\Pthree$
              greater than $|z|$ and all $p'\in\Podd$ with $-p'\in\ran(t_{s-1})$, let
              $t_s = t_{s-1}\cup \{(i,j)\mapsto -p\}$ (note that by the sufficiently large choice
              of $p$, the oracle $w_{s-1}$ is $t_s$-valid), and choose $w_s = w_{s-1}b$ with $b\in\{0,1\}$
              such that $w_s$ is $t_s$-valid.
              \\(meaning: try to ensure that $(L(M_i),L(M_j))$ is not a disjoint NP-pair. If this is
              impossible, choose a sufficiently large prime $p$ and require that the further construction
              ensures $A_p = \oli{B_p}$ and $A_p\in\cUP\cap\ccoUP$ (cf.\ V3). The treatment of tasks
              of the form $(i,j,\cdot)$ will make sure that $A_p$ cannot be reduced to $(L(M_i), L(M_j))$.)
        \item task $(i,j,r)$ with $i\ne j$: It holds $t_{s-1}(i,j) = -p$ for a prime $p\in\Pthree$,
              since otherwise, this task would have been deleted in the treatment
              of task $(i,j)$.
              Define $t_s = t_{s-1}$ and
              choose a $t_s$-valid $w_s\sqsupsetneq w_{s-1}$
              such that for some $n\in\N^+$
              one of the following two statements holds:
              \begin{itemize}
                  \item   $0^n\in A_p^{v}$ for all $v\sqsupseteq w_s$
                        and $M_i^{w_s}(T_r(0^n))$ definitely rejects.
                  \item   $0^n\in B_p^{v}$ for all $v\sqsupseteq w_s$
                        and $M_j^{w_s}(T_r(0^n))$ definitely rejects.
              \end{itemize}
              (meaning: make sure that it does not hold $(A_p, B_p) \redmprom (L(M_i),L(M_j))$ via $T_r$. As V3 ensures
              $A_p = \oli{B_p}$ relative to the final oracle, $T_r$ does not reduce $A_p$ to $(L(M_i), L(M_j))$.)
        \item task $(i,i)$: Let $t' =
                  t_{s-1}\cup \{(i,i)\mapsto 0\}$. If there
              exists a $t'$-valid partial oracle $v\sqsupsetneq w_{s-1}$,
              then let $t_s = t'$, define $w_s$ to be the least $t'$-valid partial oracle $\sqsupsetneq w_{s-1}$,
              and delete all tasks $(i,i,\cdot)$ from $T$.
              Otherwise, let $z = |w_{s-1}|$, choose some $q\in\Pone$
              greater than both $|z|$ and all $p'\in\Podd$ with $-p'\in\ran(t_{s-1})$, let
              $t_s = t_{s-1}\cup \{(i,i)\mapsto -q\}$ (note that by the sufficiently large choice of $q$, the
              oracle $w_{s-1}$ is $t_s$-valid),
              and choose $w_s = w_{s-1}b$ with $b\in\{0,1\}$
              such that $w_s$ is $t_s$-valid.
              \\(meaning: try to ensure that $M_i$ is not a $\cUP$-machine. If this is
              impossible, choose a sufficiently large prime $q\in\Pone$ and require that
              in the further construction $C_q$ remains a $\cUP$-set. The treatment of tasks of the
              form $(i,i,\cdot)$ will make sure that $C_q$ cannot be reduced to $L(M_i)$.)
        \item task $(i,i,r)$: It holds $t_{s-1}(i,i) = -q$ for a prime $q\in\Pone$,
              since otherwise, this task would have been deleted in the treatment
              of task $(i,i)$.
              Define $t_s = t_{s-1}$ and
              choose a $t_s$-valid $w_s\sqsupsetneq w_{s-1}$
              such that for some $n\in\N^+$ one of the following conditions holds:
              \begin{itemize}
                  \item   $0^n\in C_q^v$ for all $v\sqsupseteq w_s$ and
                        $M_i^{w_s}(T_r(0^n))$ definitely rejects.
                  \item   $0^n\notin C_q^v$ for all $v\sqsupseteq w_s$ and
                        $M_i^{w_s}(T_r(0^n))$ definitely accepts.
              \end{itemize}
              (meaning: ensure that $T_r$ does not reduce $C_q$ to $L(M_i)$.)
    \end{itemize}
    Observe that the choice of $t_s$ guarantees $t_s\in \calT$.
    We now show that the construction is possible.
    For that purpose, we first describe how a valid oracle can be extended
    by one bit such that it remains valid.
    \begin{claim}\label{claim_oracleextension}
        Let $s\in\N$, $w\in\az^*$ be a $t_s$-valid oracle with $w\sqsupseteq w_s$, and $z = |w|$.
        Then there exists $b\in\{0,1\}$ such that $wb$ is $t_s$-valid. In detail, the following statements hold.
        \begin{enumerate}
            \item\label{st_197400192} If $|z|$ is odd and $|z| \ne p^k$ for all $p\in\Podd$
                  with $-p\in\ran(t_s)$ and all $k\in\N^+$, then $w0$ and $w1$ are $t_s$-valid.
            \item\label{st_10932471904} If there exist $p\in\Pthree$ with $-p\in\ran(t_s)$
                  and $k\in\N^+$ such that $|z| = p^k$, $z\ne 1^{p^k}$, and
                  $w\cap \az^{p^k}=\emptyset$, then $w0$ and $w1$ are $t_s$-valid.
            \item\label{st_107924} If there exist $p\in\Pthree$ with $-p\in\ran(t_s)$ and $k\in\N^+$
                  such that $z = 1^{p^k}$ and $w\cap\az^{p^k} = \emptyset$, then $w1$ is $t_s$-valid.
            \item\label{st_14702734} If there exist $q\in\Pone$ with $-q\in\ran(t_s)$ and $k\in\N^+$
                  such that $|z| = q^k$ and $w\cap\az^{q^k}=\emptyset$, then $w0$ and $w1$ are $t_s$-valid.
            \item\label{st_178040812704} If $z = c(i,x,y)$ for $i\in\N^+$ and $x,y\in\N$, $0<t_s(i)
                      \le z$, and $F_i^w(x) = y$, then $w1$ is $t_s$-valid.
            \item\label{st_205832305} If $z = c(i,x,y)$ for $i\in\N^+$ and $x,y\in\N$, $F_i^w(x) = y\in K^w$,
                  and at least one of the three conditions (i) $t_s(i)$ undefined, (ii) $t_s(i) = 0$,
                  and (iii) $t_s(i) > z$ holds, then $w0$ and $w1$ are $t_s$-valid.
            \item\label{st_147890712034} In all other cases (i.e., none of the assumptions
                  in \ref{st_197400192}--\ref{st_205832305} holds)
                  $w0$ is $t_s$-valid.
        \end{enumerate}
    \end{claim}

    \begin{proof}
        We first show the following assertions.
        \begin{align}
            \parbox[c]{145mm}{$w0$ satisfies V1.}\label{eq_190472}                                       \\
            \parbox[c]{145mm}{If
                (i) $z = c(i,x,y)$ for $i\in\N^+$ and $x,y\in\N$ with $F_i^w(x) = y\in K^w$ or
                (ii) $z$ has odd length,
                then $w1$ satisfies V1.}\label{eq_19023472}            \\
            \parbox[c]{145mm}{$w0$ satisfies V5 unless there exist $i\in\N^+$ and $x,y\in\N$
                such that
                (i) $z = c(i,x,y)$,
                (ii) $0<t_s(i)$, (iii) $t_s(i) \le z$, and (iv) $F_i^w(x) = y$.}\label{eq_1904234342172} \\
            \parbox[c]{145mm}{$w1$ satisfies V5.}\label{eq_342566190472}
        \end{align}

        (\ref{eq_190472}) and (\ref{eq_19023472}): Let $i'\in\N^+$ and $x',y'\in\N$ such that
        $c(i',x',y')\in w$. Then, as $w$ is $t_s$-valid, by V1, $F_{i'}^w(x') = y'\in K^w$ and by
        Claim~\ref{claim_7378194873}, $F_{i'}^{w}(x')$ is definite and $y'\in K^v$ for all
        $v\sqsupseteq w$. Hence, in particular, $F_{i'}^{wb}(x') = y'\in K^{wb}$ for all $b\in
            \{0,1\}$. This shows (\ref{eq_190472}). For the proof of (\ref{eq_19023472}) it remains
        to consider $z$. In case (ii) $w1$ satisfies V1 as $|z|$ is odd and each $c(\cdot,\cdot,\cdot)$ has even length.
        Consider case (i), i.e.,
        $z = c(i,x,y)$ for $i\in\N^+$ and $x,y\in\N$ with $F_i^w(x)=y\in K^w$.
        Then by Claim~\ref{claim_7378194873}, $F_i^{w1}(x) =y\in K^{w1}$, which shows (\ref{eq_19023472}).

        (\ref{eq_1904234342172}) and (\ref{eq_342566190472}): Let $i'\in\N^+$ and $x'\in\N$
        such that $0<t_s(i')\le c(i',x',F_{i'}^w(x'))<|w|$. Then by Claim~\ref{claim_7378194873},
        $F_{i'}^w(x')$ is definite and thus, $F_{i'}^{wb}(x') = F_{i'}^w(x')$ for all $b\in\{0,1\}$.
        As $w$ is $t_s$-valid, it holds $c(i',x',F_{i'}^w(x'))\in w$ and hence,
        $c(i',x',F_{i'}^{wb}(x'))\in w \subseteq wb$ for all $b\in\{0,1\}$.
        This shows (\ref{eq_342566190472}). In order to finish the proof for (\ref{eq_1904234342172}),
        it remains to consider $z$. Assume $z = c(i,x,y)$ for some $i,x,y\in\N$ with $i>0$ (otherwise,
        $w0$ clearly satisfies V5). If (ii) or (iii) is wrong, then $w0$ satisfies V5.
        If (iv) is wrong, then $F_i^w(x)\ne y$. By Claim~\ref{claim_7378194873}, this
        computation is definite and hence, $F_i^{w0}(x)\ne y$, which is why $w0$ satisfies V5.
        This shows (\ref{eq_1904234342172}).

        \smallskip Let us now prove the assertions 1--7 and note that we do not
        have to consider V2, V4, and V6 as these conditions are not affected by
        extending a $t_s$-valid oracle.

        \begin{enumerate}[wide, labelwidth=!, labelindent=0pt,topsep=0pt,partopsep=0pt]
            \item By (\ref{eq_190472}) and (\ref{eq_19023472}), the oracles $w0$ and $w1$
                  satisfy V1.
                  By (\ref{eq_1904234342172}) and (\ref{eq_342566190472}),
                  the oracles $w0$ and $w1$ satisfy V5 (for the application of
                  (\ref{eq_1904234342172}) recall
                  that each $c(i,x,y)$ has even length and hence,
                  for all $i,x,y$ condition (i) does not hold). V3 and V7 are not affected as $|z|\ne p^k$
                  for all primes $p$ with $-p\in\ran(t_s)$ and all $k>0$.

            \item By (\ref{eq_190472}), (\ref{eq_19023472}),
                  (\ref{eq_1904234342172}), and (\ref{eq_342566190472}),
                  the oracles $w0$ and $w1$ satisfy V1 and V5.
                  As $p\in\Pthree$, V7 is satisfied by $w0$ and $w1$. Moreover,
                  $w0$ satisfies V3 as due to $z\ne 1^{p^k}$ the oracle $w0$ is not defined
                  for all words of length $p^k$. Finally, $w1$ satisfies V3 since $\az^{p^k}\cap w
                      = \emptyset$.

            \item By (\ref{eq_19023472}) and (\ref{eq_342566190472}),
                  the oracle $w1$ satisfies V1 and V5. As $p\in\Pthree$,
                  V7 is satisfied by $w1$. Moreover, as $w\cap \az^{p^k} = \emptyset$,
                  it holds $|w1\cap \az^{p^k}| = 1$ and hence, $w1$ satisfies V3.

            \item By (\ref{eq_190472}), (\ref{eq_19023472}),
                  (\ref{eq_1904234342172}), and (\ref{eq_342566190472}),
                  the oracles $w0$ and $w1$ satisfy V1 and V5. As $q\in\Pone$,
                  the oracles $w0$ and $w1$ satisfy V3. Finally, $w0$ trivially satisfies
                  V7 and $w1$ satisfies V7 as $w\cap\az^{q^k}=\emptyset$.

            \item By (\ref{eq_342566190472}), the oracle $w1$ satisfies V5.
                  As $|z|$ is even, $w1$ trivially satisfies V3 and V7.
                  It remains to argue for V1.
                  For that purpose we show $y\in K^w$.
                  Then (\ref{eq_19023472}) can be applied and $w1$ satisfies V1.
                  \\\indent By Claim~\ref{claim_7378194873},
                  $F_{i}^{w}(x)$ is definite.
                  Assume that for $z = |w|$
                  it holds $z = c(i,x,y)$ for $i\in\N^+$ and $x,y\in\N$,
                  $0<t_{s}(i)\le z$, and $F_i^w(x) = y\notin K^w$.
                  Let $s'>0$ be the step which treats the task $i$ (note $s' \le s$ as $t_s(i)$ is defined).
                  By Claim~\ref{claim_90124780576}, $w$ is $t_{s'-1}$-valid.
                  Moreover, by Claim~\ref{claim_7378194873}, $F_{i}^{w}(x) \notin K^v$
                  for all $v\sqsupseteq w$. Thus,
                  $w$ is $t'$-valid for $t' = t_{s'-1}\cup\{i\mapsto 0\}$, which is why
                  the construction would have chosen $t_{s'} = t'$, in contradiction to
                  $t_s(i) > 0$. Hence, $y\in K^w$.

            \item By (\ref{eq_190472}), (\ref{eq_19023472}), (\ref{eq_1904234342172}), and (\ref{eq_342566190472}),
                  the oracles $w0$ and $w1$ satisfy V1 and V5.
                  As $|z|$ is even, both $w0$ and $w1$ satisfy V3 and V7.

            \item By (\ref{eq_190472}), $w0$ satisfies V1. Moreover, (\ref{eq_1904234342172})
                  can be applied since otherwise, there would exist $i,x,y\in\N$ with $i>0$ such that
                  conditions (i)--(iv) of the assertion
                  (\ref{eq_1904234342172}) hold and then we were in case \ref{st_178040812704}. Hence,
                  $w0$ satisfies V5. Trivially, $w0$ satisfies V7 and finally, $w0$ satisfies V3 as
                  the only way $w0$ could hurt V3 is that $z = 1^{p^k}$ for some $p\in\Pthree$ with $-p\in\ran(t_s)$
                  and $k>0$ as well as $w \cap \az^{p^k} = \emptyset$, but this case is treated in
                  \ref{st_107924}.
        \end{enumerate}
        This finishes the proof of Claim~\ref{claim_oracleextension}.
    \end{proof}

    In order to show that the above construction is possible, assume that
    it is not possible and let $s>0$ be the least number such that step~$s$
    of the construction fails.

    If step~$s$ treats a task $\tau\in\N^+\cup(\N^+)^2$, then $t_{s-1}(\tau)$ is
    not defined, since the value of $\tau$ is defined in the
    unique treatment of the task $\tau$.  If $t_s(\tau)$ is chosen
    to be 0, then the construction clearly is possible.
    Otherwise, as was mentioned in the description of the construction,
    the choice of $t_s(\tau)$
    guarantees that the $t_{s-1}$-valid oracle $w_{s-1}$ is
    even $t_s$-valid. Then Claim~\ref{claim_oracleextension}
    ensures that there exists a $t_s$-valid $w_{s-1}b$ for some $b\in\{0,1\}$.
    Hence, the construction does not fail in step~$s$, a contradiction.

    For the remainder of the proof that the construction above is possible
    we assume that step~$s$ treats a task
    $(i,j,r)\in(\N^+)^3$. We treat
    the cases $i = j$ and $i \ne j$ simultaneously whenever it is possible.
    Recall that in the case $i = j$ we work for the diagonalization ensuring that
    $L(M_i)$ is not a complete $\cUP$-set and in the case $i\ne j$ we work for
    the diagonalization ensuring that the pair $(L(M_i),L(M_j))$ is not hard for $\cUP\cap\ccoUP$.

    In both cases, $t_s = t_{s-1}$ and $t_s(i,j)
        = -p$ for some $p\in\Podd$ (recall $p\in\Pone$ if $i = j$ and
    $p\in\Pthree$ if $i\ne j$). Let $\gamma(x) = (x^r+r)^{i+j} + i+j$
    and choose $n = p^k$ for some $k\in\N^+$
    such that \begin{equation}\label{eq_01847012}
        2^{2n-2} > 2^{n+1}\cdot \gamma(n)
    \end{equation}
    and $w_{s-1}$ is undefined for all words of length $\ge n$.
    Note that $\gamma(n)$ is not less than the running time of each
    of the computations $M_i^D(T_r(0^n))$ and $M_j^D(T_r(0^n))$
    for each oracle $D$.

    We define $u\sqsupseteq w_{s-1}$ to be the minimal $t_s$-valid partial oracle
    that is defined for all words of length $<n$. Such an oracle exists by
    Claim~\ref{claim_oracleextension}.

    Moreover, for $z\in \az^n$, let $u_z\sqsupsetneq u$
    be the minimal $t_s$-valid partial oracle with $u_z\cap\az^n = \{z\}$ that
    is defined for all words of length $\le \gamma(n)$.
    Such an oracle exists by
    Claim~\ref{claim_oracleextension}:
    first, starting from $u$ we extend the current oracle bitwise such that
    (i) it remains $t_s$-valid, (ii) it is defined for precisely the words of length $\le n$,
    and (iii) its intersection with $\az^n$ equals $\{z\}$. This is possible
    by (\ref{st_10932471904}, \ref{st_107924}, and \ref{st_147890712034}) or
    (\ref{st_14702734} and \ref{st_147890712034}) of Claim~\ref{claim_oracleextension} depending on whether
    $p\in\Pthree$ or $p\in \Pone$. Then by Claim~\ref{claim_oracleextension},
    the current oracle can be extended bitwise without losing its $t_s$-validity until it is
    defined for all words of length $\le\gamma(n)$.

    \begin{claim}\label{claim_1047120472314}
        Let $z\in\az^n$.
        \begin{enumerate}
            \item\label{st_172304231740} For each $\alpha\in u_z\cap \az^{>n}$ one of the following statements holds.
                  \begin{itemize}
                      \item   $\alpha = 1^{{p'}^\kappa}$ for some $p'\in\Pthree$ with $-p'\in\ran(t_s)$
                            and some $\kappa>0$.
                      \item   $\alpha = c(i',x,y)$ for some $i'\in\N^+$ and $x,y\in\N$ with
                            $0<t_s(i') \le c(i',x,y)$, $F_{i'}^{u_z}(x) = y$, and $y \in K^{u_z}$.
                  \end{itemize}
            \item\label{st_14790120} For all $p'\in\Pthree$
                  with $-p'\in\ran(t_s)$ and all $\kappa>0$,
                  if $n< {p'}^\kappa \le \gamma(n)$, then
                  $u_z \cap \az^{{p'}^\kappa} = \{1^{{p'}^\kappa}\}$.
        \end{enumerate}
    \end{claim}
    \begin{proof}
        \begin{enumerate}[wide, labelwidth=!, labelindent=0pt,topsep=0pt,partopsep=0pt]
            \item Let $\alpha\in u_z\cap\az^{>n}$. Moreover, let $u'$ be the prefix of $u_z$ that
                  has length $\alpha$, i.e., $\alpha$ is the least word that $u'$ is not defined for.
                  In particular, it holds $u'\cap\az^{\le n} = u_z\cap\az^{\le n}$ and thus,
                  $u'\cap\az^n = \{z\}$.
                  As $u\sqsubseteq u'\sqsubseteq u_z$ and both $u$ and $u_z$ are $t_s$-valid,
                  Claim~\ref{claim_sandwich} yields that $u'$ is also $t_s$-valid.
                  \\\indent Let us apply Claim~\ref{claim_oracleextension} to the oracle
                  $u'$. If one of the cases \ref{st_197400192}, \ref{st_10932471904},
                  \ref{st_14702734}, \ref{st_205832305}, and \ref{st_147890712034} can be applied, then $u'0$ is $t_s$-valid
                  and can be extended to a $t_s$-valid oracle $u''$ with $|u''| = |u_z|$ by
                  Claim~\ref{claim_oracleextension}. As $u''$ and $u_z$ agree on all
                  words $<\alpha$ and $\alpha\in u_z-u''$, we obtain $u'' < u_z$ and due to $u'\sqsubseteq u''$
                  we know that $u''\cap\az^n = \{z\}$. This is
                  a contradiction to the choice of $u_z$ (recall that $u_z$ is the
                  minimal $t_s$-valid partial oracle that is defined for all words of length $\le \gamma(n)$ and
                  that satisfies $u_z\cap\az^n = \{z\}$).
                  \\\indent Hence, none of the cases \ref{st_197400192}, \ref{st_10932471904},
                  \ref{st_14702734}, \ref{st_205832305}, and \ref{st_147890712034} of
                  Claim~\ref{claim_oracleextension} can be applied, i.e.,
                  either (i) $\alpha = 1^{{p'}^\kappa}$
                  for some $p'\in\Pthree$ and $\kappa>0$ with $-p'\in\ran(t_s)$ or (ii) $\alpha = c(i',x,y)$
                  for $i',x,y\in\N$, $i'>0$, and $0<t_s(i')\le \alpha$. In the latter
                  case, as $\alpha\in u_z$ and $u_z$ is $t_s$-valid, we obtain from V1 that $F_{i'}^{u_z}(x) = y\in K^{u_z}$.

            \item
                  As $-p'\in\ran(t_s)$, $u_{z}$ is $t_s$-valid, and $u_z$ is defined
                  for all words of length $p'^\kappa$, V3 yields that
                  there exists $\beta\in\az^{{p'}^\kappa}\cap u_{z}$.
                  Let $\beta$ be the minimal element of $\az^{{p'}^\kappa}\cap u_{z}$.
                  It suffices to show $\beta = 1^{{p'}^\kappa}$.
                  For a contradiction, we assume $\beta < 1^{{p'}^\kappa}$.
                  Let $u'$ be the prefix of $u_z$ that is defined for exactly the words
                  of length $< {p'}^\kappa$. Then $u\sqsubseteq u'\sqsubseteq u_z$ and
                  both $u$ and $u_z$ are $t_s$-valid. Hence, by Claim~\ref{claim_sandwich},
                  the oracle $u'$ is $t_s$-valid as well.
                  \\\indent By Claim~\ref{claim_oracleextension}, $u'$ can be extended
                  to a $t_s$-valid oracle $u''$ that satisfies $|u''| = |u_z|$
                  and $u''\cap\az^{{p'}^\kappa} = \{ 1^{{p'}^\kappa}\}$.
                  Then $\beta\in u_{z}-u''$. As the oracles $u''$ and $u_{z}$ agree on all words smaller than $\beta$,
                  we have $u'' < u_{z}$ and $u''\cap\az^n = \{z\}$, in contradiction
                  to the choice of $u_{z}$ (again, recall that $u_z$ is the
                  minimal $t_s$-valid partial oracle that is defined for all words of length $\le \gamma(n)$ and
                  that satisfies $u_z\cap\az^n = \{z\}$).
        \end{enumerate}
        This finishes the proof of Claim~\ref{claim_1047120472314}.
    \end{proof}

    Let us study the case that for some odd (resp., even) $z\in\az^n$
    the computation $M_i^{u_z}(T_r(0^n))$ (resp., $M_j^{u_z}(T_r(0^n))$ if $z$ is even)
    rejects. Then it even definitely rejects since $u_z$ is defined
    for all words of length $\le\gamma(n)$.
    If $i\ne j$, then $p\in\Pthree$ and since $z\in u_z$, we have
    $0^n\in A_p^{v}$ for all $v\sqsupseteq u_z$ (resp., $0^n\in B_p^v$
    for all $v\sqsupseteq u_z$ if $z$ is even).
    Analogously, if $i = j$, then $p\in \Pone$ and as $z\in u_z$, we have
    $0^n\in C_p^v$ for all $v\sqsupseteq u_z$.
    Hence, in all these cases we can choose $w_s = u_z$ and obtain a contradiction
    to the assumption that step~$s$ of the construction fails in treating the task $(i,j,r)$.
    Therefore, for the remainder of the proof that the construction is possible
    we assume the following:
    \begin{itemize}\label{assumption_1}
        \item   For each odd $z\in\az^n$ the computation
              $M_i^{u_z}(T_r(0^n))$ definitely accepts.
        \item   For each even $z\in\az^n$ the computation
              $M_j^{u_z}(T_r(0^n))$ definitely accepts.
    \end{itemize}
    Note that in case $i = j$ we could have also formulated the two conditions
    equivalently in the following simpler way: for each $z\in\az^n$ the computation $M_i^{u_z}(T_r(0^n))$
    definitely accepts. Recall, however, that as far as possible we consider the cases $i = j$ and $i\ne j$
    simultaneously.

    We will show that this assumption leads to a contradiction, which will ensure
    that the above construction is possible.

    Let $U_z$ for $z\in\az^n$ odd (resp., $z\in\az^n$ even) be the set of all oracle
    queries of the least accepting path of $M_i^{u_z}(T_r(0^n))$ (resp.,
    $M_j^{u_z}(T_r(0^n))$). Observe $\ell(U_z) \le \gamma(n)$.
    Moreover, define $Q_0(U_z) = U_z$ and for $m\in\N$,
    \begin{align*}
        Q_{m+1}(U_z) = & \bigcup_{\substack{c(i',x,y)\in Q_m(U_z)\\
                                 i',x,y\in\N, i'>0}} \Big[\{q\mid
        \text{$q$ is queried by $F_{i'}^{u_z}(x)$}\}\cup          \\
                       & \{q\mid
        \parbox[t]{110mm}{$y = \pairing{ {0}^{i''}, 0^{|x'|^{i''}+i''}, x'}$ for some $i''>0$ and $x'\in\N$,
            $M_{i''}^{u_z}(x')$ has an accepting path, and $q$ is queried by the least such path$\}\Big]$.}
    \end{align*}
    Let $Q(U_z) = \bigcup_{m \in\N} Q_m(U_z)$.  Note that for $c(i',x,y)\in Q_m(U_z)$ for some
    $m$ it does not necessarily hold $y\in K^{u_z}$ and therefore,
    the computation $M_{i''}^{u_z}(x')$
    (in the notation used in the equation above) might not have any accepting paths.
    In that case the second of the two sets in the equation above is empty.
    \begin{claim}\label{claim_2140771289032}
        For all $z\in\az^n$, $\ell(Q(U_z)) \le 2 \ell(U_z) \le 2\gamma(n)$ and
        the length of each word in $Q(U_z)$ is $\le \gamma(n)$.
    \end{claim}
    \begin{proof}
        We show that for all $m\in\N$,
        $\ell(Q_{m+1}(U_z))\le \nicefrac{1}{2}\cdot \ell(Q_m(U_z))$. Then
        $\sum_{m = 0}^\kappa \nicefrac{1}{2^m} \le 2$ for all $\kappa\in\N$
        implies $\ell(Q(U_z))\le 2\cdot\ell(U_z)$. Moreover,
        the second part of the claim follows from
        $\ell(U_z)\le \gamma(n)$ and $\ell(Q_{m+1}(U_z))\le \nicefrac{1}{2}\cdot \ell(Q_m(U_z))$.

        Let $m\in \N$ and consider an arbitrary element $\alpha$ of $Q_m(U)$. If $\alpha$
        is not of the form $c(i',x,y)$ for $i'\in\N^+$ and $x,y\in\N$,
        then $\alpha$ generates no elements in $Q_{m+1}(U)$.
        Assume $\alpha = c(i',x,y)$ for $i'\in\N^+$ and $x,y\in \N$.
        The computation $F_{i'}^v(x)$ runs for at most
        $|x|^{i'}+i' < \nicefrac{|\alpha|}{4}$ steps, where \eqq{$<$} holds
        by (\ref{eq_140712041003}).
        Hence, the set of queries $Q$ of $F_{i'}^{v}(x)$ satisfies
        $\ell(Q)\le\nicefrac{|\alpha|}{4}$.

        Let us assume $y=\pairing{ {0}^{i''}, 0^{|x'|^{i''}+i''}, x'}$ for $i''\in\N^+$ and $x'\in\N$
        (otherwise, the second of the two sets in the definition of $Q_{m+1}(U_z)$ is empty and does
        not require further consideration).
        Then the computation $M_{i''}^v(x)$ runs for less than
        $|x'|^{i''}+i'' < |y| < \nicefrac{|\alpha|}{4}$ steps, where again ``$<$'' holds by (\ref{eq_140712041003}).
        Hence, for the set $Q$ of queries
        of the least accepting path of the computation $M_{i''}^{v}(x)$ (if such a path exists) we have
        $\ell(Q) \le \nicefrac{|\alpha|}{4}$.
        Consequently,
        \begin{align*}\ell(Q_{m+1}(U)) & \le  \sum_{\substack{c(i',x,y)\in Q_m(U_z)\\
                                                      i',x,y\in\N, i'>0}}
                      \Big[\underbrace{\ell\big(\{q\mid
                               \text{$q$ is queried by $F_{i'}^{u_z}(x)$}\}\big)}_{
                               \le \nicefrac{|c(i',x,y)|}{4}} +
                      \\&
                      \phantom{\le \sum_{\substack{c(i',x,y)\in Q_m(U_z)\\
                                  i',x,y\in\N, i'>0}}}  \underbrace{\ell\big(\{q\mid
                                                            \parbox[t]{90mm}{$y = \pairing{ {0}^{i''}, 0^{|x'|^{i''}+i''}, x'}$ for some $i''>0$ and $x'\in\N$,
                                                                $M_{i''}^{u_z}(x')$ has an accepting path, and $q$ is queried by the least such path$\}\big)\Big]$}}_{
                                                            \le \nicefrac{|c(i',x,y)|}{4}} \\
                                       & \le \sum_{\substack{c(i',x,y)\in Q_m(U_z)\\
                                                     i',x,y\in\N, i'>0}} \nicefrac{|c(i',x,y)|}{2}
                      \\&\le \nicefrac{\ell(Q_m(U_z))}{2},
        \end{align*}
        which finishes the proof of Claim~\ref{claim_2140771289032}.
    \end{proof}

    We now define a similar notion. Let $Q_0'(U_z) = U_z$ and for $m\in\N$,
    \begin{align*}
        Q_{m+1}'(U_z) = & \bigcup_{\substack{c(i',x,y)\in Q_m'(U_z)\\
                                  i',x,y\in\N, i'>0}} \{q\mid
        \text{$q$ is queried by $F_{i'}^{u_z}(x)$}\}.
    \end{align*}
    Moreover, define $Q'(U_z) = \bigcup_{m \in\N} Q_m'(U_z)$.
    By definition $Q'_m(U_z)\subseteq Q_m(U_z)$ for all $m\in\N$ and hence,
    $Q'(U_z) \subseteq Q(U_z) \subseteq \az^{\le\gamma(n)}$, where the latter
    ``$\subseteq$'' holds by Claim~\ref{claim_2140771289032}.

    For $z,z'\in\az^n$ we say that $z$ and $z'$ {\em conflict} (resp., {\em strongly conflict})
    if there is a word $\alpha\in Q(U_z)\cap Q(U_{z'})$ (resp.,  $\alpha\in Q'(U_z)\cap Q'(U_{z'})$)
    which is in $u_z\triangle u_{z'}$. In that case, we say
    $z$ and $z'$ (strongly) conflict in $\alpha$.
    Note that whenever $z$ and $z'$ (strongly) conflict in
    a word $\alpha$, then $|\alpha|\ge n$,
    as $u_z$ and $u_{z'}$ are both extensions of $u$
    and $u$ is defined for all words of length $< n$. By definition,
    if $z$ and $z'$ strongly conflict,
    then $z$ and $z'$ conflict.

    The next four claims are dedicated to the purpose of proving that for
    each odd $z\in {\Sigma }^{n}$ and each even $z'\in {\Sigma }^{n}$, it holds that
    $z$ and $z'$ conflict in a word of length $n$.
    Indeed, then $z$ and $z'$ conflict in one of the words
    $z$ and $z'$ as these are the only words of length $n$ in $u_{z}
        \cup u_{z'}$. Once we have proven this, we will be able to generate a
    contradiction that completes the proof that the oracle construction
    described above is possible.
    \begin{claim}\label{claim_0197412}
        Let $z,z'\in\az^n$ such that $z$ is odd and $z'$ is even.
        If $z$ and $z'$ conflict, then they conflict
        in a word of length $n$.
    \end{claim}
    \begin{proof}
        Let $\alpha$ be the least word in which $z$ and $z'$ conflict
        (note that $|\alpha|\le \gamma(n)$ due to $\alpha\in Q(U_z)\cap Q(U_{z'})$ and
        Claim~\ref{claim_2140771289032}).
        Then $\alpha\in u_z\triangle u_{z'}$.
        By symmetry, it suffices to consider the case $\alpha\in u_z-u_{z'}$.
        For a contradiction, assume that $|\alpha|\ne n$, which implies $|\alpha|>n$
        as $u_z\sqsupseteq u$, $u_{z'}\sqsupseteq u$, and $u$ is defined for
        all words of length $<n$. Then by Claim~\ref{claim_1047120472314}, two situations
        are possible.

        \begin{enumerate}[wide, labelwidth=!, labelindent=0pt]
            \item If $\alpha = 1^{{p'}^\kappa}$ for $p'\in\Pthree$ with
                  $-p'\in\ran(t_s)$ and $\kappa>0$, then by Claim~\ref{claim_1047120472314}.\ref{st_14790120},
                  $\alpha\in u_{z'}$, a contradiction.

            \item Here, $\alpha = c(i',x,y)$ for $i'\in\N^+$
                  and $x,y\in\N$ with $0<t_s(i') \le c(i',x,y)$
                  and $F_{i'}^{u_z}(x) = y \in K^{u_z}$.
                  Then $F_{i'}^{u_{z'}}(x)\ne y$, since otherwise, by the $t_s$-validity
                  of $u_{z'}$ and V5, it would hold $\alpha\in u_{z'}$.
                  Consequently, $F_{i'}^{u_{z'}}(x) \ne F_{i'}^{u_z}(x)$.
                  Hence, there exists a query $\beta$ that is asked by both $F_{i'}^{u_z}(x)$ and
                  $F_{i'}^{u_{z'}}(x)$
                  and that is in $u_z\triangle u_{z'}$ (otherwise, both computations would
                  output the same word). By definition of $Q(U_z)$ and $Q(U_{z'})$, it
                  holds $\beta\in Q(U_z)\cap Q(U_{z'})$. Hence, $z$ and $z'$
                  conflict in $\beta$ and $|\beta| \le |x|^{i'}+i' < |c(i',x,y)| = |\alpha|$,
                  in contradiction to the assumption that $\alpha$ is the least word which
                  $z$ and $z'$ conflict in.
        \end{enumerate}
        In both cases we obtain a contradiction. Thus, the proof is complete.
    \end{proof}

    We want to show next that for all
    odd $z\in\az^n$ and all even $z'\in\az^n$ it holds that
    $z$ and $z'$ indeed conflict.

    Recall that $u\sqsupseteq w_{s-1}$ is the minimal $t_s$-valid partial oracle
    that is defined for all words of length $<n$. Let $z\in\az^n$. We say that
    a partial oracle $v\sqsupseteq u$ is {\it consistent} with $u_z$ relative to $Q(U_z)$ (resp., relative to $Q'(U_z)$)
    if $v(q) = u_z(q)$ for all $q\in Q(U_z)$ (resp., all $q\in Q'(U_z)$)
    that $v$ is defined for (i.e., $q<|v|)$. Note that since $v$ and $u_z$ are
    both extensions of $u$, it trivially holds $v(q) = u_z(q)$ for all $q\in\az^{<n}$.
    As an abbreviation, whenever we say that $v$ is consistent with $u_z$, then we mean that
    $v$ is consistent with $u_z$ {\it relative to $Q(U_z)$.}

    \begin{claim}\label{claim_1847010237} The following statements hold.
        \begin{enumerate}
            \item \label{st_1742014897}
                  Let $t=t_{s'}$ for some $0\le s' \le s$ and $z,z'\in\az^n$ such that $z$ and
                  $z'$ do not conflict.
                  For each $t$-valid partial oracle $v\sqsupsetneq u$ that is defined for exactly the words of length $\le n$
                  and consistent with both $u_z$ and $u_{z'}$, there exists a $t$-valid partial oracle $v'\sqsupsetneq v$
                  that is consistent with both $u_z$ and $u_{z'}$ and satisfies $|v'| = |u_z|$.
            \item  \label{st_1498612894}Let $z\in\az^n$.
                  For each $t_s$-valid partial oracle $v\sqsupsetneq u$ that is defined for exactly the words of length $\le n$
                  and consistent with $u_z$ relative to $Q'(U_z)$,
                  there exists a $t_s$-valid partial oracle $v'\sqsupsetneq v$ that is
                  consistent with $u_z$ relative to $Q'(U_z)$ and satisfies $|v'| = |u_z|$.
        \end{enumerate}
    \end{claim}
    Note that in the notation of the claim above, $v$ is consistent with
    $u_z$ if and only if for all $q\in Q(U_z)\cap\az^n$,
    it holds $v(q) = 1$ if $q = z$ and $v(q) = 0$ otherwise. Moreover, recall that
    $|v'| = |u_z|$ expresses that $v'\text{ is defined for exactly the words of length }\le\gamma(n)$.

    \medskip \begin{proof}{\bf of Claim~\ref{claim_1847010237}}
        The two statements can be proven in a very similar way.
        Basically, the proof of the second statement
        is a simplified version of the proof of the first statement. Nevertheless,
        for the sake of completeness, we also give a separate proof for the second
        statement.
        \begin{enumerate}[wide, labelwidth=!,labelindent=0pt]\item
                  Let $w\sqsupseteq v$ with $|w| < |u_z|$ be $t$-valid and consistent with
                  $u_z$ and $u_{z'}$.
                  Moreover, let $\alpha = |w|$, i.e., $\alpha$ is the least word
                  that $w$ is not defined for. It suffices
                  to show the following:
                  \begin{enumerate}
                      \item If $\alpha = 0^{{p'}^\kappa}$ for a $p'\in\Pthree$
                            with $-p'\in\ran(t_s)$ and $\kappa>0$, then there exists a $t$-valid
                            $w'\sqsupsetneq w$ that is defined for exactly the words of length $\le {p'}^\kappa$
                            and consistent with $u_z$ and $u_{z'}$.

                            Note that in this case $|w'|\le |u_z|$ since $u_z$ is
                            defined for exactly the words of length $\le \gamma(n)$.
                      \item If for all $p'\in\Pthree$
                            with $-p'\in\ran(t_s)$ and all $\kappa >0$ the word $\alpha$ is not of
                            length ${p'}^\kappa$, then there exists $b\in\{0,1\}$ such that
                            $wb$ is $t$-valid and consistent with $u_z$ and $u_{z'}$.
                  \end{enumerate}

                  \begin{enumerate}
                      \item Assume $\alpha = 0^{{p'}^\kappa}$ for some $p'\in\Pthree$
                            with $-p'\in\ran(t_s)$ and $\kappa>0$. Then we let $w'\sqsupsetneq w$ be the minimal partial oracle
                            that is defined for all words of length $\le {p'}^\kappa$ and contains $1^{{p'}^\kappa}$, i.e.,
                            $w' = w\cup \{1^{{p'}^\kappa}\}$ when interpreting the partial oracles as sets.
                            As $u_z\cap \az^{{p'}^\kappa} =  u_{z'} \cap \az^{{p'}^\kappa} = \{1^{{p'}^\kappa}\}$ by
                            Claim~\ref{claim_1047120472314}.\ref{st_14790120} (note $|\alpha|>n$ since $\alpha =|w|$,
                            $w\sqsupseteq v$, and $v$ is defined for all words of length $\le n$), we obtain that $w'$
                            is consistent with $u_z$ and $u_{z'}$. Moreover, if $-p'\in\tn{ran}(t)$, then $w'$ is $t$-valid
                            by Claim~\ref{claim_oracleextension}.\ref{st_10932471904}
                            and Claim~\ref{claim_oracleextension}.\ref{st_107924}. If $-p'\notin \tn{ran}(t)$, then
                            $w'$ is $t$-valid by Claim~\ref{claim_oracleextension}.\ref{st_197400192}.

                      \item Here we study two subcases.
                            \begin{enumerate}[wide, labelwidth=!, labelindent=0pt,topsep=0pt,partopsep=0pt]
                                \item Assume that $\alpha=c(i',x,y)$ for
                                      $i'\in\N^+$ and $x,y\in\N$ with $0<t_s(i') \le \alpha$.
                                      Let us first assume that $\alpha\notin Q(U_z)\cup Q(U_{z'})$. Then there exists $b\in\{0,1\}$
                                      such that $wb$ is $t$-valid (cf.\ Claim~\ref{claim_oracleextension})
                                      and clearly $wb$ is consistent with $u_z$ and $u_{z'}$.

                                      From now on we assume $\alpha\in Q(U_z)\cup Q(U_{z'})$. By symmetry, it suffices to
                                      consider the case $\alpha\in Q(U_z)$. As all queries $q$
                                      of $F_{i'}^{u_z}(x)$ are in $Q(U_z)$ and due to $|q|\le |x|^{i'} + i' < \alpha$ satisfy
                                      $u_z(q) = w(q)$, it holds \begin{equation}F_{i'}^w(x) = F_{i'}^{u_z}(x).\label{eq_14876091865012}\end{equation} We study two cases.
                                      \begin{enumerate}
                                          \item   If $\alpha\in u_z$, then by V1, $F_{i'}^{u_z}(x) = y\in K^{u_z}$. By (\ref{eq_14876091865012}),
                                                $F_{i'}^w(x) = y$. Let us show $y\in K^w$:
                                                As $y\in K^{u_z}$, it holds $y = \pairing{ {0}^{i''},0^{|x'|^{i''} + i''}, x'}$ for some $i''>0$
                                                and $x'\in\N$, the computation $M_{i''}^{u_z}(x')$ has an accepting path, and all
                                                queries $q$ of the least accepting path of this computation are in $Q(U_z)$ and due to
                                                $|q|\le |y| < |\alpha|$ satisfy $u_z(q) = w(q)$. Hence, $M_{i''}^w(x')$ accepts
                                                and $y\in K^w$. Let us choose $b = 1$. Note that $t(i')$ is not necessarily defined.
                                                If $t(i')$ is defined, then $0 < t(i') = t_s(i')\le \alpha$, we apply
                                                Claim~\ref{claim_oracleextension}.\ref{st_178040812704}, and obtain that $wb$ is
                                                $t$-valid. If $t(i')$ is undefined, then we apply
                                                Claim~\ref{claim_oracleextension}.\ref{st_205832305} and obtain that $wb$ is
                                                $t$-valid. Clearly $wb$ is consistent with $u_z$. In order to see
                                                that $wb$ is consistent with $u_{z'}$, it suffices
                                                to show $\big(\alpha\in Q(U_{z'})\Rightarrow \alpha\in u_{z'}\big)$. This holds since
                                                otherwise, $z$ and $z'$ conflict.

                                          \item   Assume $\alpha\notin u_z$. Then by V5, $F_{i'}^{u_z}(x) \ne y$. By (\ref{eq_14876091865012}),
                                                $F_{i'}^w(x) \ne y$. Choose $b = 0$. Then by
                                                Claim~\ref{claim_oracleextension}.\ref{st_147890712034}, $wb$ is $t$-valid and clearly
                                                $wb$ is consistent with $u_z$. In order to see that $wb$ is consistent with $u_{z'}$,
                                                it suffices to argue for $\alpha$. If $\alpha\in Q(U_{z'})$, then
                                                $\alpha\notin u_{z'}$ as otherwise, $z$ and $z'$ would conflict.
                                      \end{enumerate}

                                \item We now consider the remaining cases, i.e., we may assume
                                      \begin{itemize}
                                          \item $\alpha$ is not of length
                                                ${p'}^\kappa$ for all $p'\in\Pthree$ with $-p'\in\ran(t_s)$ and all $\kappa>0$ and
                                          \item $\alpha \ne c(i',x,y)$ for all
                                                $i'\in\N^+$ and $x,y\in\N$ with $0<t_s(i') \le \alpha$.
                                      \end{itemize}
                                      In this case, it holds
                                      $\alpha\notin u_z\cup u_{z'}$ by Claim~\ref{claim_1047120472314}.\ref{st_172304231740}.
                                      We choose $b = 0$ and obtain that $wb$ is consistent with $u_z$ and $u_{z'}$. Moreover,
                                      by Claim~\ref{claim_oracleextension}, $wb$ is $t$-valid ($w0$ is $t$-valid unless
                                      Claim~\ref{claim_oracleextension}.\ref{st_107924} or Claim~\ref{claim_oracleextension}.\ref{st_178040812704}
                                      is applicable, which is ---by assumption--- not the case).
                            \end{enumerate}\end{enumerate}

            \item Let $w\sqsupseteq v$ with $|w| < |u_z|$ be $t_s$-valid and consistent with
                  $u_z$ relative to $Q'(U_z)$.
                  Moreover, let $\alpha = |w|$. It suffices
                  to show the following:
                  \begin{enumerate}
                      \item If $\alpha = 0^{{p'}^\kappa}$ for a $p'\in\Pthree$
                            with $-p'\in\ran(t_s)$ and $\kappa>0$, then there exists a $t_s$-valid
                            $w'\sqsupsetneq w$ that is defined for exactly the words of length $\le {p'}^\kappa$
                            and consistent with $u_z$ relative to $Q'(U_z)$.

                            Note that in this case $|w'|\le |u_z|$ since $u_z$ is
                            defined for exactly the words of length $\le \gamma(n)$.
                      \item If for all $p'\in\Pthree$
                            with $-p'\in\ran(t_s)$ and all $\kappa >0$ the word $\alpha$ is not of
                            length ${p'}^\kappa$, then there exists $b\in\{0,1\}$ such that
                            $wb$ is $t_s$-valid and consistent with $u_z$ relative to $Q'(U_z)$.
                  \end{enumerate}

                  \begin{enumerate}
                      \item Assume $\alpha = 0^{{p'}^\kappa}$ for some $p'\in\Pthree$
                            with $-p'\in\ran(t_s)$ and $\kappa>0$. Then we let $w'\sqsupsetneq w$ be the minimal partial oracle
                            that is defined for all words of length $\le{p'}^\kappa$ and contains $1^{{p'}^\kappa}$, i.e.,
                            $w' = w\cup \{1^{{p'}^\kappa}\}$ when interpreting the partial oracles as sets.
                            As $u_z\cap \az^{{p'}^\kappa} = \{1^{{p'}^\kappa}\}$ by
                            Claim~\ref{claim_1047120472314}.\ref{st_14790120} (note $|\alpha|>n$ since $\alpha =|w|$,
                            $w\sqsupseteq v$, and $v$ is defined for all words of length $\le n$), we obtain that $w'$
                            is consistent with $u_z$ relative to $Q'(U_z)$. Moreover, $w'$ is $t_s$-valid
                            by Claim~\ref{claim_oracleextension}.\ref{st_10932471904}
                            and Claim~\ref{claim_oracleextension}.\ref{st_107924}.

                      \item Here we study two subcases.
                            \begin{enumerate}[wide, labelwidth=!, labelindent=0pt,topsep=0pt,partopsep=0pt]
                                \item Assume that $\alpha=c(i',x,y)$ for
                                      $i'\in\N^+$ and $x,y\in\N$ with $0<t_s(i') \le \alpha$.
                                      Let us first assume that $\alpha\notin Q'(U_z)$. Then there exists $b\in\{0,1\}$
                                      such that $wb$ is $t_s$-valid (cf.\ Claim~\ref{claim_oracleextension})
                                      and clearly $wb$ is consistent with $u_z$ relative to $Q'(U_z)$.

                                      From now on we assume $\alpha\in Q'(U_z)$.  As all queries $q$
                                      of $F_{i'}^{u_z}(x)$ are in $Q'(U_z)$ and due to $|q|\le |x|^{i'} + i' < \alpha$ satisfy
                                      $u_z(q) = w(q)$, it holds \begin{equation}F_{i'}^w(x) = F_{i'}^{u_z}(x).\label{eq_58234969}\end{equation} We study two cases.
                                      If $\alpha\in u_z$, then by V1, $F_{i'}^{u_z}(x) = y\in K^{u_z}$. By (\ref{eq_58234969}),
                                      $F_{i'}^w(x) = y$.  Let us choose $b = 1$. Applying
                                      Claim~\ref{claim_oracleextension}.\ref{st_178040812704}, we obtain that $wb$ is
                                      $t$-valid. Clearly $wb$ is consistent with $u_z$ relative to $Q'(U_z)$.

                                      Assume $\alpha\notin u_z$. Then by V5, $F_{i'}^{u_z}(x) \ne y$. By (\ref{eq_58234969}),
                                      $F_{i'}^w(x) \ne y$. Choose $b = 0$. Then by
                                      Claim~\ref{claim_oracleextension}.\ref{st_147890712034}, $wb$ is $t_s$-valid and clearly
                                      $wb$ is consistent with $u_z$ relative to $Q'(U_z)$.

                                \item We now consider the remaining cases, i.e., we may assume
                                      \begin{itemize}
                                          \item $\alpha$ is not of length
                                                ${p'}^\kappa$ for all $p'\in\Pthree$ with $-p'\in\ran(t_s)$ and all $\kappa>0$ and
                                          \item $\alpha \ne c(i',x,y)$ for all
                                                $i'\in\N^+$ and $x,y\in\N$ with $0<t_s(i') \le \alpha$.
                                      \end{itemize}
                                      In this case, it holds
                                      $\alpha\notin u_z$ by Claim~\ref{claim_1047120472314}.\ref{st_172304231740}.
                                      We choose $b = 0$ and obtain that $wb$ is consistent with $u_z$ relative to $Q'(U_z)$. Moreover,
                                      by Claim~\ref{claim_oracleextension}, $wb$ is $t_s$-valid ($w0$ is $t_s$-valid unless
                                      Claim~\ref{claim_oracleextension}.\ref{st_107924} or Claim~\ref{claim_oracleextension}.\ref{st_178040812704}
                                      is applicable, which is ---by assumption--- not the case).
                            \end{enumerate}\end{enumerate}\end{enumerate}
        This finishes the proof of Claim~\ref{claim_1847010237}.
    \end{proof}
    \begin{claim}\label{claim_1470174012451}
        For all $z\in\az^n$, it holds $z\in Q'(U_z)$.
    \end{claim}
    \begin{proof}
        For a contradiction,
        assume $z\notin Q'(U_z)$ for some $z\in\az^n$.
        We study the cases $i = j$ and $i\ne j$ separately.

        First let us consider the case $i \ne j$.
        Here, $p\in\Pthree$.
        By symmetry, it suffices to consider the case
        that $z$ is odd. Let $z'$
        be the minimal even element of $\az^n$ that is not in $Q'(U_z)$.
        Such $z'$ exists as $2^{n-1}> 4\gamma(n)>2\gamma(n)$ by (\ref{eq_01847012}),
        $\ell(Q'(U_z))\le \ell(Q(U_z)) \le 2\gamma(n)$ by Claim~\ref{claim_2140771289032},
        and hence, $|Q'(U_z)|\le \ell(Q'(U_z)) \le 2\gamma(n) < 2^{n-1} =
            |\{z''\in\az^n\mid z''\tn{ even}\}|$. Now choose $u'$
        to be the partial oracle that is defined for exactly the words of length $\le n$
        and that satisfies $u' = u\cup\{z'\}$ when the partial oracles are considered as sets. Then
        $u'$ is $t_s$-valid by Claim~\ref{claim_oracleextension}.\ref{st_10932471904},
        Claim~\ref{claim_oracleextension}.\ref{st_107924},
        and Claim~\ref{claim_oracleextension}.\ref{st_147890712034}.
        Moreover, as $z,z'\notin Q'(U_z)$,
        the oracle $u'$ is consistent with $u_z$ relative to $Q'(U_z)$.
        Hence, we can apply
        Claim~\ref{claim_1847010237}.\ref{st_1498612894} to the oracle $u'$ for the parameter $z$ and
        obtain a $t_s$-valid partial oracle $v$ that is
        defined for all words of length $\le \gamma(n)$,
        satisfies $v\cap\az^n =\{z'\}$, and is consistent with $u_z$ relative to $Q'(U_z)$.
        Hence, $v(q) = u_z(q)$ for
        all $q\in Q'(U_z)$. By this property and by the fact that $U_z\subseteq Q'(U_z)$ contains all queries
        asked by the least accepting path of $M_i^{u_z}(T_r(0^n))$ (such a path exists by the assumptions
        made on page~\pageref{assumption_1}), this path is also an accepting path of the computation
        $M_i^v(T_r(0^n))$. As $v$ is defined for all words
        of length $\le \gamma(n)$, the computation
        $M_i^v(T_r(0^n))$ definitely accepts.
        Let us study two cases depending on whether $M_j^v(T_r(0^n))$ definitely
        accepts or definitely rejects
        (note that this computation is definite as $v$ is
        defined for all words of length $\le \gamma(n)$):
        \begin{itemize}
            \item Assume that $M_j^v(T_r(0^n))$ definitely accepts. Let $s'>0$ be the step
                  that treats the task $(i,j)$. Hence, $s' \le s$ since $t_s(i,j)$ is defined.
                  By Claim~\ref{claim_90124780576}, the oracle $v$
                  is $t_{s'-1}$-valid. Now,
                  as both $M_i^v(T_r(0^n))$ and $M_j^v(T_r(0^n))$ definitely accept, $v$ is
                  even $t'$-valid for $t' = t_{s'-1}\cup\{(i,j)\mapsto 0\}$. But then the construction
                  would have chosen $t_{s'} = t'$,
                  in contradiction to $t_s(i,j) \ne 0$.
            \item Assume that $M_j^v(T_r(0^n))$ definitely rejects.
                  As $v\cap\az^n = \{z'\}$, it holds  $0^n\in B_p^{v'}$
                  for all $v'\sqsupseteq v$. This is a contradiction
                  to the assumption that step~$s$
                  of the construction fails.
        \end{itemize}
        As in both cases we obtain a contradiction, the proof of the first statement
        is complete.

        \bigskip Now assume $i = j$. Here, $p\in\Pone$.
        Let $u'$ be the partial oracle that is defined for exactly the words of length $\le n$
        and satisfies $u' = u$ when the partial oracles are considered as sets.
        Then $u'$ is $t_s$-valid by
        Claim~\ref{claim_oracleextension}.\ref{st_14702734} and $u'$ is consistent with $u_z$ relative to $Q'(U_z)$
        as both oracles are extensions of $u$, $u_z\cap\az^n = \{z\}$, and $z\notin Q'(U_z)$.
        Thus, we can apply Claim~\ref{claim_1847010237}.\ref{st_1498612894} to the oracle $u'$. Hence,
        there exists a $t_s$-valid partial oracle $v$ that
        satisfies $|v| = |u_z|$, $v\cap\az^n = \emptyset$,
        and that is consistent with $u_z$ relative to $Q'(U_z)$. The first condition expresses that $v$
        is defined for all words of length $\le\gamma(n)$, in particular for
        all words in $Q'(U_z)$). Thus, as $v$ is consistent with $u_z$, it holds
        $v(q) = u_z(q)$ for all $q\in Q'(U_z)$.
        By this property and the fact that $U_z\subseteq Q'(U_z)$ contains all queries
        asked by the least accepting path of $M_i^{u_z}(T_r(0^n))$ (such a path exists by the assumptions
        made on page~\pageref{assumption_1}), this
        path is also an accepting path of the computation
        $M_i^v(T_r(0^n))$. As $v$ is defined for all words of length $\le \gamma(n)$, the computation
        $M_i^v(T_r(0^n))$ is definite.
        Thus, $0^n\notin C_q^{v'}$ for all $v'\sqsupseteq v$ and
        $M_i^v(T_r(0^n))$ definitely accepts, in contradiction to the assumption that step~$s$
        of the construction fails.
    \end{proof}

    \begin{claim}\label{claim_01247102372122}
        For all odd $z\in\az^n$ and all even $z'\in\az^n$, it holds that $z$ and $z'$ conflict.
    \end{claim}
    \begin{proof}Assume there are $z$ odd and $z'$ even such that $z$ and $z'$
        do not conflict. Then let $u'\sqsupsetneq u$ be the minimal partial oracle that is defined for
        all words of length $\le n$ and contains $z$ and $z'$, i.e., interpreting partial oracles as sets
        it holds $u' = u \cup \{z,z'\}$. Let $s'>0$ be the step that treats the task $(i,j)$. Then
        $s'\le s$ as $t_s(i,j)$ is defined. As $t_s\in\calT$ is injective on its support and $t_s(i,j) = -p$,
        it holds $-p\notin\ran(t_{s'-1})$. Therefore,
        $u'$ is $t_{s'-1}$-valid by Claim~\ref{claim_oracleextension}.\ref{st_197400192}
        (recall that $u$ is $t_{s'-1}$-valid by Claim~\ref{claim_90124780576}).
        In order to apply Claim~\ref{claim_1847010237}.\ref{st_1742014897} to $u'$ for
        the parameters $z$, $z'$, and $s'-1$, it is sufficient to show that
        $u'$ is consistent with $u_z$ and $u_{z'}$. Assume this is not the case. Then due to
        $u_z\cap\az^n= \{z\}$ and $u_{z'}\cap\az^n = \{z'\}$ it holds
        $z\in Q(U_{z'})$ or $z'\in Q(U_z)$. As by Claim~\ref{claim_1470174012451},
        $z\in Q'(U_z)\subseteq Q(U_z)$ and $z'\in Q'(U_{z'}) \subseteq Q(U_{z'})$,
        one of the words $z$ and $z'$ is in $Q(U_z)\cap Q(U_{z'})$.
        Since $u_z$ and $u_{z'}$ disagree on both $z$ and $z'$, it holds that
        $z$ and $z'$ conflict, a contradiction. Hence,
        Claim~\ref{claim_1847010237}.\ref{st_1742014897} can be applied to
        $u'$ for the parameters $z$, $z'$, and $s'-1$.

        Applying Claim~\ref{claim_1847010237}.\ref{st_1742014897}, we obtain a $t_{s'-1}$-valid
        $v\sqsupseteq u'$ that is defined for all words of
        length $\le \gamma(n)$ and is consistent with $u_z$ and $u_{z'}$, i.e.,
        $v(q) = u_z(q)$ for all $q\in Q(U_z)$ and $v(q) = u_{z'}(q)$ for all $q\in Q(U_{z'})$
        (recall $Q(U_z)\cup Q(U_{z'})\subseteq \az^{\le\gamma(n)}$, which is why $v$ is
        defined for all words in $Q(U_z)\cup Q(U_{z'})$).
        We claim
        \begin{equation}\label{eq_12704271890341}
            \parbox[c]{145mm}{$v$ is $t'$-valid for $t' = t_{s'-1}\cup\{(i,j)\mapsto 0\}$.}
        \end{equation}
        Once (\ref{eq_12704271890341}) is proven, we obtain a contradiction
        as then the construction would have chosen $t_{s'} = t'$, in contradiction
        to $t_s(i,j) \ne 0$. Hence, then our assumption is
        wrong and for all odd $z\in\az^n$ and all even $z'\in\az^n$, it holds that
        $z$ and $z'$ conflict.

        It remains to prove (\ref{eq_12704271890341}). We study two cases.

            {\bf Case 1}: first we assume $i\ne j$. In this case
        it suffices to prove that $M_i^v(T_r(0^n))$ and
        $M_j^v(T_r(0^n))$ definitely accept.
        Recall that $M_i^{u_z}(T_r(0^n))$ and $M_j^{u_{z'}}(T_r(0^n))$
        definitely accept, $v(q) = u_z(q)$ for all $q\in
            Q(U_z)$, and $v(q) = u_{z'}(q)$ for all $q\in Q(U_{z'})$.
        This implies that the least accepting paths of $M_i^{u_z}(T_r(0^n))$
        and $M_j^{u_{z'}}(T_r(0^n))$ are also accepting paths of the computations
        $M_i^v(T_r(0^n))$ and $M_j^v(T_r(0^n))$. Moreover, the latter computations
        are definite by the choice of $v$. Thus, $v$ is $t'$-valid.

            {\bf Case 2}: assume $i = j$. We have to prove that on some input
        $x$ the computation $M_i^v(x)$ has two accepting paths. By Claim~\ref{claim_1470174012451},
        $z\in Q'(U_z)$ and $z'\in Q'(U_{z'})$. As $z$ and $z'$ do not
        conflict, $z$ and $z'$ do not strongly conflict and
        it holds $z\notin Q'(U_{z'})$, which implies $Q'(U_z)\ne Q'(U_{z'})$. Let $\kappa\in\N$
        be minimal such that $Q_\kappa'(U_z)\ne Q_\kappa'(U_{z'})$ and for a contradiction, assume $\kappa>0$.

        Let $\alpha \in Q_\kappa'(U_z)\triangle Q_\kappa'(U_{z'})$. Without loss of generality, we assume
        $\alpha \in Q_\kappa'(U_z) - Q_\kappa'(U_{z'})$. Then there exist
        $i',x,y\in\N$ with $i'>0$ such that $c(i',x,y)\in Q_{\kappa-1}(U_z)$
        and $F_{i'}^{u_z}(x)$ asks the query $\alpha$.
        By the choice of $\kappa$, it holds $Q_{\kappa-1}'(U_{z'}) = Q_{\kappa-1}'(U_z)$
        and thus, $c(i',x,y)\in Q_{\kappa-1}'(U_{z'})$.
        Hence, all queries of $F_{i'}^{u_{z'}}(x)$
        are in $Q_\kappa(U_{z'})$. However, $\alpha\notin Q_\kappa(U_{z'})$ and therefore,
        $\alpha$ is not asked by $F_{i'}^{u_{z'}}(x)$. This shows that there is a word
        $\beta\in u_z\triangle u_{z'}$ asked by both $F_{i'}^{u_z}(x)$ and $F_{i'}^{u_{z'}}(x)$
        (otherwise, the two computations would ask the same queries).
        But then $\beta \in Q_\kappa'(U_z)\cap Q_\kappa'(U_{z'})$, which implies that $z$
        and $z'$ strongly conflict, a contradiction to the assumption that $z$ and $z'$ do not conflict.
        Hence, $\kappa = 0$ and $U_z=Q_0'(U_z)\ne Q_0'(U_{z'}) = U_{z'}$.

        Recall that $U_z$ (resp., $U_{z'}$) is the set consisting of all
        oracle queries of the least accepting path $P$ (resp., $P'$) of the computation
        $M_i^{u_z}(T_r(0^n))$ (resp., $M_i^{u_{z'}}(T_r(0^n))$). As
        $u_z(q) = v(q)$ for all $q\in  Q(U_z)\supseteq U_z$ and $u_{z'}(q) = v(q)$
        for all $q\in Q(U_{z'})\supseteq U_{z'}$, the paths $P$ and $P'$ are accepting
        paths of the computation $M_i^v(T_r(0^n))$. Finally, $P$ and $P'$ are distinct
        paths since $U_z$ and $U_{z'}$ are distinct sets (also recall that the transducer does not query the oracle).
        This finishes the proof of (\ref{eq_12704271890341})
        and the proof of Claim~\ref{claim_01247102372122}.
    \end{proof}
    The remainder of the proof that the construction is possible is based on an
    idea by Hartmanis and Hemachandra \cite{hh88}.
    Consider the set
    \begin{align}
        E & = \{\{z,z'\}\mid z,z'\in\az^n, \text{$z$ odd $\Leftrightarrow z'$ even},
        (z\in Q(U_{z'}) \vee z'\in Q(U_z))\}\nonumber                                                              \\
          & \subseteq \bigcup_{z\in\Sigma^n} \{\{z,z'\}\mid z'\in\Sigma^n, z'\in Q(U_z) \}.\label{eq_189472193421}
    \end{align}
    Let $z,z'\in\az^n$ such that $(z\tn{ odd}\Leftrightarrow z'\tn{ even})$. Then by
    Claim~\ref{claim_01247102372122} and Claim~\ref{claim_0197412},
    $z$ and $z'$ conflict in a word of length $n$. As $u_z \triangle u_{z'} =
        \{z,z'\}$, this means that they conflict in $z$ or $z'$. Hence, $z\in Q(U_{z'})$ or
    $z'\in Q(U_z)$.

    This shows $E = \{\{z,z'\}\mid z,z'\in\az^n, \text{$z$ odd
    $\Leftrightarrow z'$ even}\}$ and thus, $|E| = 2^{2n-2}$. By Claim~\ref{claim_2140771289032},
    for each $z\in\az^n$ it holds $|Q(U_z)|\le \ell(Q(U_z))\le 2 \gamma(n)$. Consequently,
    $$|E| \stackrel{(\ref{eq_189472193421})}{\le}
        \sum_{z\in\az^n} |Q(U_z)| \le 2^n\cdot 2 \gamma(n) = 2^{n+1}\cdot \gamma(n)
        \stackrel{(\ref{eq_01847012})}{<} 2^{2n-2} = |E|,$$ a contradiction. Hence,
    the assumption that the construction fails in step~$s$ treating the
    task $(i,j,r)$ is wrong. This shows that the construction described above is possible and
    $O$ is well-defined. In order to finish the proof of Theorem~\ref{theorem_0917240914},
    it remains to show that
    \begin{itemize}
        \item $\DisjNP^O$ does not contain a pair $\redm[,O]$-hard for $\cUP^O\cap\ccoUP^O$,
        \item each non-empty problem in $\cNP^O$ has a $\cP^O$-optimal proof system, and
        \item $\cUP^O$ does not contain a $\redm[,O]$-complete problem.
    \end{itemize}

    \begin{claim}\label{claim_10749109234971}
        $\DisjNP^O$ does not contain a pair that is $\redm[,O]$-hard
        for $\cUP^O\cap\ccoUP^O$.
    \end{claim}
    \begin{proof}
        Assume the assertion is wrong, i.e., there exists a pair in $\DisjNP^O$ that is
        $\redm[,O]$-hard for $\cUP^O\cap\ccoUP^O$.
        By Proposition~\ref{prop:queryfree}, there also exists a pair in $\DisjNP^O$ that is
        $\redm$-hard for $\cUP^O\cap\ccoUP^O$.
        Hence, there exist distinct $i,j\in\N^+$ such that $(L(M_i^O),L(M_j^O)) \in \DisjNP^O$
        and for every $A\in\cUP^O\cap\ccoUP^O$ it holds $A\redm (L(M_i^O),L(M_j^O))$.
        From $L(M_i^O) \cap L(M_j^O) = \emptyset$ it follows that
        for all $s$ there does not exist $z$ such that both
        $M_i^{w_s}(z)$ and $M_j^{w_s}(z)$ definitely accept.
        Hence, by V2, there does not exist $s$ with $t_s(i,j) = 0$ and thus, by construction
        $t_s(i,j) = -p$ for some $p \in \Pthree$ and all sufficiently large $s$.
        The latter implies $|O\cap \az^{p^k}| = 1$ for all $k>0$ (cf.\ V3),
        which yields $A_p^O = \oli{B_p^O}$ and $A_p^O\in\cUP^O\cap\ccoUP^O$.
        Since $(L(M_i^O),L(M_j^O))$ is $\redm$-hard for $\cUP^O\cap\ccoUP^O$, there exists $r$ such that
        $A_p^O \redm (L(M_i^O),L(M_j^O))$ via $T_r$.
        Let $s$ be the step that treats task $(i,j,r)$.
        This step makes sure that there exists $n\in\N^+$ such that
        at least one of the following properties holds:
        \begin{itemize}
            \item   $0^n\in A_p^{v}$ for all $v\sqsupseteq w_{s}$
                  and $M_i^{w_{s}}(T_r(0^n))$ definitely rejects.
            \item   $0^n\in B_p^{v}$ for all $v\sqsupseteq w_{s}$
                  and $M_j^{w_{s}}(T_r(0^n))$ definitely rejects.
        \end{itemize}
        As $O(q) = w_s(q)$ for all $q$ that $w_s$ is defined for, one of the
        following two statements holds.
        \begin{itemize}
            \item $0^n \in A_p^O$ and $T_r(0^n)$ is rejected by $M_i^O$.
            \item $0^n \in B_p^O = \oli{A_p^O}$ and $T_r(0^n)$ is rejected by $M_j^O$.
        \end{itemize}
        This is a contradiction to $A_p^O \redm (L(M_i^O),L(M_j^O))$ via $T_r$,
        which completes the proof of Claim~\ref{claim_10749109234971}.
    \end{proof}

    \begin{claim}\label{claim_289017410824}
        Each non-empty problem in $\cNP^O$ has a $\cP^O$-optimal proof system.
    \end{claim}
    \begin{proof}
        By Corollary~\ref{coro_pps_oracle}, it suffices to prove that $K^O$
        has a $\cP^O$-optimal proof system.

        Let $g\in \FP^O$ be an arbitrary proof system for $K^O$ and
        $a$ be an arbitrary element of $K^O$.
        Define $f$ to be the following function $\sow\to\sow$:
        $$
            f(z) = \begin{cases}
            g(z')&\text{if $z = 1z'$}\\
            y&\text{if $z = 0c(i,x,y)$ for $i\in\N^+$, $x,y\in\N$, and $c(i,x,y)\in O$}\\
            a&\text{otherwise}
            \end{cases}
        $$
        By definition, $f\in\FP^O$ and as $g$ is a proof system for $K^O$ it holds
        $f(\sow)\supseteq K^O$. We show $f(\sow)\subseteq K^O$. Let $z\in\sow$. Assume $z=0c(i,x,y)$
        for $i\in\N^+$, $x,y\in\N$, and $c(i,x,y)\in O$ (otherwise, clearly $f(z)\in K^O$).
        Let $s$ be large enough such that $w_s$ is defined for $c(i,x,y)$, i.e., $w_s(c(i,x,y)) =1$.
        As $w_s$ is $t_s$-valid, we obtain by V1 that $y\in K^{w_s}$ and by
        Claim~\ref{claim_7378194873} that $y\in K^{v}$ for all $v\sqsupseteq w_s$.
        Thus, $y\in K^O$,
        which shows that $f$ is a proof system for $K^O$.

        It remains to show that each proof system for $K^O$ is $\cP^O$-simulated by $f$.
        Let $h$ be an arbitrary proof system for $K^O$. Then there exists $i>0$ such that
        $F_i^O$ computes $h$. By construction,
        $t_s(i) > 0$, where $s$ is the number of the step that treats the task $i$.
        Consider the following function
        $\pi:\sow\to\sow$:
        $$
            \pi(x) = \begin{cases}
            0c(i,x,F_i^O(x))&\text{if $c(i,x,F_i^O(x))\ge t_s(i)$}\\
            z&\text{if $c(i,x,F_i^O(x)) < t_s(i)$ and $z$ is minimal with $f(z) = F_i^O(x)$}
            \end{cases}
        $$
        As $f$ and $F_i^O$ are proof systems for $K^O$, for every $x$ there exists $z$ with
        $f(z) = F_i^O(x)$. Hence, $\pi$ is total.
        Since $t_s(i)$ is a constant, $\pi\in \FP^O$. It remains to show that
        $f(\pi(x)) = F_i^O(x)$ for all $x\in\sow$. If $c(i,x,F_i^O(x)) < t_s(i)$, it holds $f(\pi(x)) = F_i^O(x)$.
        Otherwise, choose $s'$ large enough such that (i) $t_{s'}(i)$ is defined (i.e., $t_{s'}(i) = t_s(i)>0$) and
        (ii) $w_{s'}$ is defined for $c(i,x,F_i^{w_{s'}}(x))$. Then, as $w_{s'}$ is $t_{s'}$-valid,
        V5 yields that $c(i,x,F_i^{w_{s'}}(x))\in w_{s'}$.
        By Claim~\ref{claim_7378194873}, $F_i^{w_{s'}}(x)$ is definite and hence,
        $F_i^O(x) = F_i^{w_{s'}}(x)$ as well as $c(i,x,F_i^O(x))\in w_{s'}\subseteq O$.
        Hence, $f(\pi(x)) = F_i^O(x)$,
        which shows $h=F_i^O\psim[,O] f$.
        This completes the proof of Claim~\ref{claim_289017410824}.
    \end{proof}

    \begin{claim}\label{claim_108472018471}
        $\cUP^O$ does not contain a $\redm[,O]$-complete problem.
    \end{claim}
    \begin{proof}
        Assume the assertion is wrong, i.e., $\cUP^O$ has a $\redm[,O]$-complete problem.
        By Proposition~\ref{prop:queryfree}, $\cUP^O$ also has a $\redm$-complete problem.
        Hence, there exists $i>0$ such that $L(M_i^O)$ is $\redm$-complete for $\cUP^O$.
        As $L(M_i^O) \in \cUP^O$, on every input $M_i^O$ has at most one accepting path.
        Thus, there does not exist $s>0$ with $t_s(i,i) = 0$ (cf.\ V6).
        Hence, by construction $t_s(i,i) = -q$ for some $q\in\Pone$ and all sufficiently large $s$.
        Then $|O\cap\az^{q^k}|\le 1$ for all $k>0$ (cf.\ V7) and consequently, $C_q^O\in\cUP^O$.
        Since $L(M_i^O)$ is $\redm$-complete for $\cUP^O$, there exists $r>0$ such that
        $C_q^O\redm L(M_i^O)$ via $T_r$.
        Let $s>0$ be the step that treats the task $(i,i,r)$. By construction,
        there exists $n\in\N^+$ such that one of the following two statements holds:
        \begin{itemize}
            \item   $0^n\in C_q^v$ for all $v\sqsupseteq w_s$ and
                  $M_i^{w_s}(T_r(0^n))$ definitely rejects.
            \item   $0^n\notin C_q^v$ for all $v\sqsupseteq w_s$ and
                  $M_i^{w_s}(T_r(0^n))$ definitely accepts.
        \end{itemize}
        As $O$ and $w_s$ agree on all words that $w_s$ is defined for, one of the following
        two conditions holds:
        \begin{itemize}
            \item   $0^n\in C_q^O$ and $M_i^O(T_r(0^n))$ rejects.
            \item   $0^n\notin C_q^O$ and $M_i^O(T_r(0^n))$ accepts.
        \end{itemize}
        This is a contradiction to $C_q^O\redm L(M_i^O)$ via $T_r$, which shows that $\cUP^O$
        does not have $\redm[,O]$-complete problems. This completes the proof of
        Claim~\ref{claim_108472018471}.
    \end{proof}
    Now the proof of Theorem~\ref{theorem_0917240914} is complete.
\end{proof}

\section*{Acknowledgement} The author thanks the anonymous referee
for many helpful comments and especially for finding a gap in the proof of the
main theorem in an earlier version of this article.
Furthermore, he thanks Fabian Egidy for a hint that helped fixing another gap in an earlier
version of this paper.

\bibliographystyle{alpha}
\bibliography{mybib}



\end{document}

%% file: comphead007.tex
\newcommand{\opstyle}[1]{\mathrm{#1}}

\newcommand{\redstyle}[1]{\mathnormal{#1}}

\newcommand{\mr}{\:\!}
\newcommand{\ml}{\:\:\!\!\!}

\newcommand{\cP}{\opstyle{P}}
\newcommand{\cNP}{\opstyle{NP}}
\newcommand{\ccoNP}{\opstyle{coNP}}
\newcommand{\cUP}{\opstyle{UP}}
\newcommand{\ccoUP}{\opstyle{coUP}}

\newcommand{\FP}{\opstyle{FP}}

\newcommand{\Pol}{\opstyle{Pol}}

\newcommand{\DisjNP}{\opstyle{DisjNP}}

\newcommand{\calC}{{\cal C}}

\newcommand{\calT}{{\cal T}}

\newcommand{\calV}{{\cal V}}
\newcommand{\calW}{{\cal W}}

\newcommand{\reduction}[3][]{%
    \redstyle{\le_{\mathrm{#3}}^{\mathrm{#2}#1}}}
\newcommand{\polyreduction}[2][]{\reduction[#1]{p}{#2}}
\newcommand{\redm}[1][]{\polyreduction[#1]{m}}

\newcommand{\redprom}[2][]{\reduction[#1]{pp}{#2}}
\newcommand{\redmprom}[1][]{\redprom[#1]{m}}

\newtheorem{dummytheorem}{Dummy-Theorem}[section]
\newcommand{\proofendsign}{$\Box$} 
\newtheorem{definition}[dummytheorem]{Definition}
\newtheorem{lemma}[dummytheorem]{Lemma}
\newtheorem{theorem}[dummytheorem]{Theorem}

\newtheorem{proposition}[dummytheorem]{Proposition}
\newtheorem{corollary}[dummytheorem]{Corollary}

\newtheorem{claim}[dummytheorem]{Claim}

\newenvironment{proof}{{\noindent \bf Proof }}{%
    {\hspace*{\fill}\proofendsign\par\bigskip}}
\newenvironment{beweis}{{\noindent \bf Beweis }}{%
    {\hspace*{\fill}\proofendsign\par\bigskip}}

\newcommand{\lqq}{\lq\lq}   
\newcommand{\rqq}{\rq\rq}   
\newcommand{\elqq}{\lqq}    
\newcommand{\erqq}{\rqq}    

\newcommand{\eqq}[1]{\elqq #1\erqq}

\newcommand{\oli}[1]{\overline{#1}}

\newcommand{\az}{{\Sigma}} 
\newcommand{\sow}{{\az^*}}

\newcommand{\rem}[1]{}

\newcommand{\tn}[1]{\textnormal{#1}}    

\newsavebox{\einzugbox}

\newenvironment{lind}[1]{\begin{indentation}{#1}{0mm}}{\end{indentation}}

\newlength{\chrlengtha}
\newlength{\chrlengthb}

\newcommand{\algob}{\begin{lind}{5mm}\fns}
\newcommand{\algoe}{\end{lind}}

\newcommand{\N}{\mathbb{N}}
\newcommand{\Z}{\mathbb{Z}}

\newcommand{\pairing}[1]{\langle #1 \rangle}

\newcommand{\fns}{\footnotesize}

%% file: paper.bbl
\begin{thebibliography}{GSSZ04}

\bibitem[BKM09]{bkm09}
O.~Beyersdorff, J.~K{\"{o}}bler, and J.~Messner.
\newblock Nondeterministic functions and the existence of optimal proof systems.
\newblock {\em Theor. Comput. Sci.}, 410(38-40):3839--3855, 2009.

\bibitem[CR79]{cr79}
S.~Cook and R.~Reckhow.
\newblock The relative efficiency of propositional proof systems.
\newblock {\em Journal of Symbolic Logic}, 44:36--50, 1979.

\bibitem[DG19]{dg19}
T.~Dose and C.~Gla{\ss}er.
\newblock {NP}-completeness, proof systems, and disjoint {NP}-pairs.
\newblock Technical Report 19-050, Electronic Colloquium on Computational Complexity {(ECCC)}, 2019.

\bibitem[Dos19a]{dos19a}
T.~Dose.
\newblock {\dosescomment{b}$\mathrm{P}\ne\mathrm{NP}$ and all sets in $\mathrm{NP}\cup\mathrm{coNP}$ have P-optimal proof systems relative to an oracle}.
\newblock {\em arXiv e-prints}, page arXiv:1909.02839, Sep 2019.

\bibitem[Dos19b]{dos19b}
T.~Dose.
\newblock {\dosescomment{c}$\mathrm{P}$-optimal proof systems for each set in $\mathrm{coNP}$ and no complete problems in $\mathrm{NP}\cap\mathrm{coNP}$ relative to an oracle}.
\newblock {\em arXiv e-prints}, page arXiv:1910.08571, Oct 2019.

\bibitem[ESY84]{esy84}
S.~Even, A.~L. Selman, and J.~Yacobi.
\newblock The complexity of promise problems with applications to public-key cryptography.
\newblock {\em Information and Control}, 61:159--173, 1984.

\bibitem[EY80]{ey80}
S.~Even and Y.~Yacobi.
\newblock Cryptocomplexity and {NP}-completeness.
\newblock In {\em Proceedings 7th International Colloquium on Automata, Languages and Programming}, volume~85 of {\em Lecture Notes in Computer Science}, pages 195--207. Springer, 1980.

\bibitem[GS88]{gs88}
J.~Grollmann and A.~L. Selman.
\newblock Complexity measures for public-key cryptosystems.
\newblock {\em SIAM Journal on Computing}, 17(2):309--335, 1988.

\bibitem[GSSZ04]{gssz04}
C.~Gla{\ss}er, A.~L. Selman, S.~Sengupta, and L.~Zhang.
\newblock Disjoint {NP}-pairs.
\newblock {\em SIAM Journal on Computing}, 33(6):1369--1416, 2004.

\bibitem[HH88]{hh88}
J.~Hartmanis and L.~A. Hemachandra.
\newblock Complexity classes without machines: On complete languages for {UP}.
\newblock {\em Theor. Comput. Sci.}, 58:129--142, 1988.

\bibitem[Kha19]{kha19}
E.~Khaniki.
\newblock {New relations and separations of conjectures about incompleteness in the finite domain}.
\newblock {\em arXiv e-prints}, pages 1--25, Apr 2019.

\bibitem[KM00]{km00}
J.~K{\"o}bler and J.~Messner.
\newblock Is the standard proof system for sat p-optimal?
\newblock In S.~Kapoor and S.~Prasad, editors, {\em FSTTCS 2000: Foundations of Software Technology and Theoretical Computer Science}, pages 361--372, Berlin, Heidelberg, 2000. Springer Berlin Heidelberg.

\bibitem[KMT03]{kmt03}
J.~K{\"o}bler, J.~Messner, and J.~Tor\'an.
\newblock Optimal proof systems imply complete sets for promise classes.
\newblock {\em Information and Computation}, 184(1):71--92, 2003.

\bibitem[KP89]{kp89}
J.~Kraj{\'{\i}}{\v c}ek and P.~Pudl{\'a}k.
\newblock Propositional proof systems, the consistency of first order theories and the complexity of computations.
\newblock {\em Journal of Symbolic Logic}, 54:1063--1079, 1989.

\bibitem[MP91]{mp91}
N.~Megiddo and C.~H. Papadimitriou.
\newblock On total functions, existence theorems and computational complexity.
\newblock {\em Theor. Comput. Sci.}, 81(2):317--324, 1991.

\bibitem[Pap94]{pap94}
C.~M. Papadimitriou.
\newblock {\em {Computational complexity}}.
\newblock Addison-Wesley, Reading, Massachusetts, 1994.

\bibitem[Pud96]{pud96}
P.~Pudl{\'a}k.
\newblock On the lengths of proofs of consistency.
\newblock In {\em Collegium Logicum}, pages 65--86. Springer Vienna, 1996.

\bibitem[Pud13]{pud13}
P.~Pudl{\'{a}}k.
\newblock {\em Logical Foundations of Mathematics and Computational Complexity - {A} Gentle Introduction}.
\newblock Springer monographs in mathematics. Springer, 2013.

\bibitem[Pud17]{pud17}
P.~Pudl{\'a}k.
\newblock Incompleteness in the finite domain.
\newblock {\em The Bulletin of Symbolic Logic}, 23(4):405--441, 2017.

\bibitem[Raz94]{raz94}
A.~A. Razborov.
\newblock On provably disjoint {NP}-pairs.
\newblock {\em Electronic Colloquium on Computational Complexity {(ECCC)}}, 1(6), 1994.

\end{thebibliography}
